\documentclass[twocolumn,preprintnumbers,amsmath,amssymb,floatfix,prd,nofootinbib]{revtex4}
\pdfoutput=1
\usepackage{epsfig}
\usepackage{dcolumn}
\usepackage{color}

\begin{document}
\title{Semi-inclusive Deep-Inelastic Scattering, Parton Distributions and \\ 
Fragmentation Functions at a Future Electron-Ion Collider}
\author{Elke C. Aschenauer}
\email{elke@bnl.gov}
\affiliation{Physics Department, Brookhaven National Laboratory, Upton, NY 11973, USA
}
\author{Ignacio Borsa}
\email{iborsa@df.uba.ar}
\affiliation{Departamento de F\'{\i}sica and IFIBA,  Facultad de Ciencias Exactas y Naturales, 
Universidad de Buenos Aires, Ciudad Universitaria, Pabell\'on\ 1 (1428) Buenos Aires, Argentina}
\author{Rodolfo Sassot}
\email{sassot@df.uba.ar} 
\affiliation{Departamento de F\'{\i}sica and IFIBA,  Facultad de Ciencias Exactas y Naturales, 
Universidad de Buenos Aires, Ciudad Universitaria, Pabell\'on\ 1 (1428) Buenos Aires, Argentina}
\author{Charlotte Van Hulse}
\email{cvhulse@mail.desy.de}
\affiliation{University of the Basque Country – UPV/EHU, Spain}
\affiliation{School of Physics, University College Dublin, Dublin, Ireland}
\begin{abstract}

We present a quantitative assessment of the impact a future Electron-Ion Collider would 
have in the determination of parton distribution functions in the proton and parton-to-hadron 
fragmentation functions through semi-inclusive deep-inelastic electron-proton scattering data. 
Specifically, we estimate the kinematic regions for which the forthcoming data are 
expected to have the most significant impact in the precision of these distributions, 
computing the respective correlation and sensitivity coefficients. Using a reweighting 
technique for the sets of simulated data with their realistic uncertainties for two different 
center-of-mass energies, we analyse the resulting new sets of parton distribution functions and 
fragmentation functions, which have significantly reduced uncertainties.

\end{abstract}
%
%
\maketitle

\section{Introduction and Motivation}
%
The quest for a quantitative picture of lepton-hadron and hadron-hadron interactions in 
terms of the basic constituents of matter and in the framework of perturbative Quantum 
Chromodynamics (pQCD) involves non-perturbative quantities that encode the 
details about the internal structure of hadrons and the mechanism leading to confinement. 
Parton distribution functions (PDFs) \cite{Butterworth:2015oua} and fragmentation functions 
(FFs) \cite{Metz:2016swz} 
stand out among these essential ingredients needed for a theoretical description of hard 
scattering processes. In the last two decades remarkable progress has been made to determine 
these non-perturbative inputs, but the need for calculations of hadronic processes with 
unprecedented precision, to validate the Standard Model of fundamental
interactions and our picture of matter at extreme conditions, gives 
the improvement of our knowledge of PDFs and FFs a crucial role in the searches for 
new physical phenomena. 

The requirement for increased precision becomes especially relevant in the case of quarks
generated through QCD radiation (\textit{sea quarks}), which are typically less constrained than their valence 
counterparts, due to the comparatively reduced flavour separation power of the data generally 
included in global analyses \cite{Butterworth:2015oua,Rojo:2015acz,Rojo:2016ymp}. An appealing
solution to this lack of stringent constraints for the sea quark distributions is to take
advantage of data from  hadron production in semi-inclusive deep-inelastic scattering
(SIDIS), which probe different quark flavour combinations depending on the final-state hadron.
The idea, originally proposed by Feynman and Field \cite{Feynman:1973xc,Field:1976ve},
has never been exploited in modern global PDF extractions since it involves, on the one
hand, the cumbersome task of a simultaneous PDF and FF extraction \cite{Borsa:2017vwy},
and on the other hand, it requires access to semi-inclusive data, 
of the same precision as the inclusive data. 
While recent semi-inclusive data \cite{Lees:2013rqd,Leitgab:2013qh,Airapetian:2012ki,Adolph:2016bwc,Agakishiev:2011dc,Abelev:2014laa} have been important to reduce 
the uncertainties on the fragmentation functions, the precision of these extractions is 
still behind compared to the one achieved for valence quark PDFs, due to the higher 
statistical precision and the kinematic coverage of totally inclusive data.

A US-based Electron Ion collider (EIC) \cite{Accardi:2012qut,Aschenauer:2014cki} 
with high-energy and high-luminosity, capable 
of a versatile range of beam energies, polarisations, and ion species will, for the first
time, be able to systematically explore and map out the dynamical
system that is the ordinary QCD bound state. The EIC is foreseen to play a transformative 
role in the understanding of the rich variety of structures at the subatomic scale.
It will open up the unique opportunity to go far beyond the present one-dimensional 
picture of nuclei and nucleons at high energy, where the composite nucleon appears as a bunch of 
fast-moving (anti-)quarks and gluons of which transverse momenta or spatial extent
are not resolved. Specifically, by correlating the information of the quark and gluon 
longitudinal momentum component with their transverse momentum and spatial distribution inside the
nucleon, it will enable nuclear “femtography”.  Such femtographic images will provide, for
the first time, insight into the QCD dynamics inside hadrons, such as the interplay between
sea quarks and gluons.
The EIC's landmark in precision and kinematic coverage for SIDIS processes will provide 
differential and accurate constraints on the distributions that quantify the structure of 
the proton and of nuclei, and on their counterparts in the final-state that describe the 
fragmentation of quarks and gluons into hadrons \cite{Aschenauer:2015ata,Aschenauer:2016our}.
In particular, the EIC will allow to probe unprecedentedly low ranges in longitudinal parton 
momentum fraction in SIDIS, over various decades in photon virtuality squared, hereby allowing 
to probe sea quarks for the first time with very high precision.

In this paper we are assessing the impact that future EIC charged 
pion and kaon SIDIS data would have on PDFs and FFs, with particular focus on sea-quark 
distributions and the possibility to see charge and flavour symmetry breaking among them. 
In order to quantify that impact, we follow the strategy discussed in \cite{Borsa:2017vwy}, 
but now using EIC pseudodata with realistic uncertainties. 

The approach relies heavily on the application of the so-called reweighting technique for 
PDFs and FFs, developed by the NNPDF collaboration \cite{Ball:2010gb,Ball:2011gg} and extended 
to a Hessian uncertainty analysis \cite{Paukkunen:2014zia}. The method allows to modify PDFs 
or FFs in order to incorporate the information coming from data sets that were not included 
in their original global extractions, avoiding a full time-consuming refit, but preserving 
the statistical rigor for the uncertainty estimates. The method has already been successfully 
demonstrated in different applications 
\cite{Ball:2010gb,Ball:2011gg,Armesto:2013kqa,Paukkunen:2014zia}. 
Another useful tool to assess the impact of new data in a global fit is to define and 
calculate correlation and sensitivity coefficients between the experimental data under 
consideration and PDFs or FFs. These also give a comparative estimate of the impact in 
different kinematic regions \cite{Wang:2018heo,Bertone:2018ecm}.

Using the above mentioned tools, we have found that the forthcoming EIC pion and kaon SIDIS 
data will have a significant impact in the determination of PDFs
and FFs not only for sea quarks but also for the up and down quark 
distributions in the proton and the favoured FFs for pion and kaons.
The improvement in the parton distributions is most noticeable for the strange 
quarks, specially for values of the Bjorken variable $x_{B}$ below $10^{-2}$, which are 
comparatively less determined in modern PDF fits.  Our results also highlight the advantage 
a high center-of-mass (c.m.s.) energy configuration of the EIC could have in 
the determination of the PDFs, as well as in constraining charge and flavour symmetry breaking 
among the proton constituents, due to the extended reach to lower $x_{B}$, which can in leading order (LO) 
be associated with the momentum fraction of the incoming nucleon taken by the struck quark in the electron 
rest frame. 

The remainder of the paper is organised as follows: in the next sections we briefly 
comment on the next-to-leading order (NLO) theoretical description of SIDIS, and on the generation of the EIC 
pseudodata and the corresponding uncertainties for the different energy configurations 
under consideration. Then, we sketch very briefly the main features of the PDF reweighting 
technique, its extension to FFs evaluated within the Hessian approach, and how it applies 
to the present study. In Section \ref{subsec:correlation}, we present the results for the 
correlation and sensitivity coefficient calculations, assessing the kinematic region where 
the new data is expected to constrain PDFs and FFs most. In Section \ref{subsec:reweighting} 
we discuss in detail the outcome of our reweighting exercise for the EIC pseudodata, with 
special interest in the light sea quark distributions and possible flavour and charge 
symmetry breaking. We briefly summarise the main results and conclusions in Section 
\ref{sec:summary}.

\begin{figure*}[t]\centering
  \begin{minipage}{0.48\textwidth}\centering
    \epsfig{file=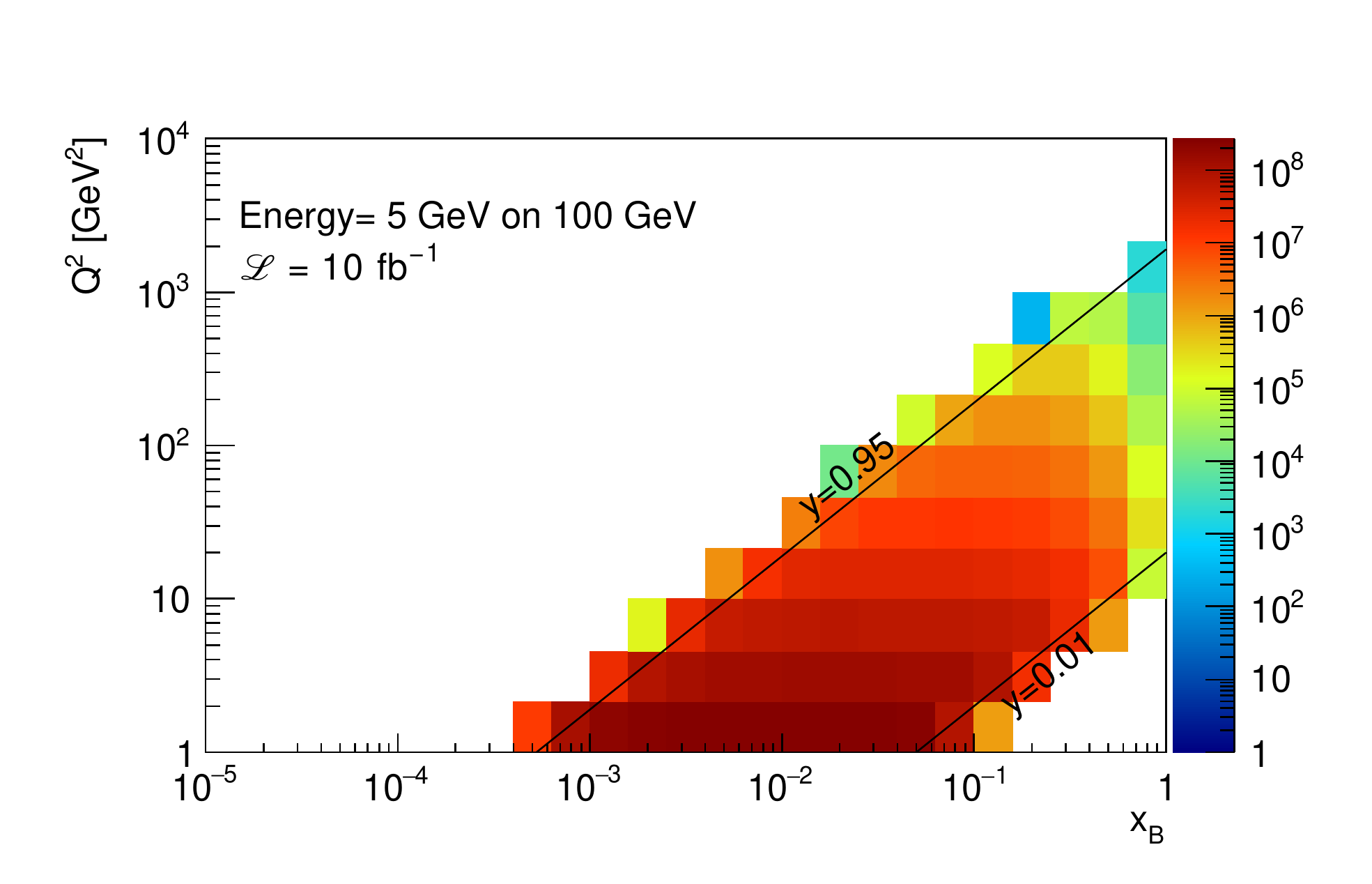,width=1.0\textwidth}
  \end{minipage}
  \begin{minipage}{0.48\textwidth}\centering
    \epsfig{file=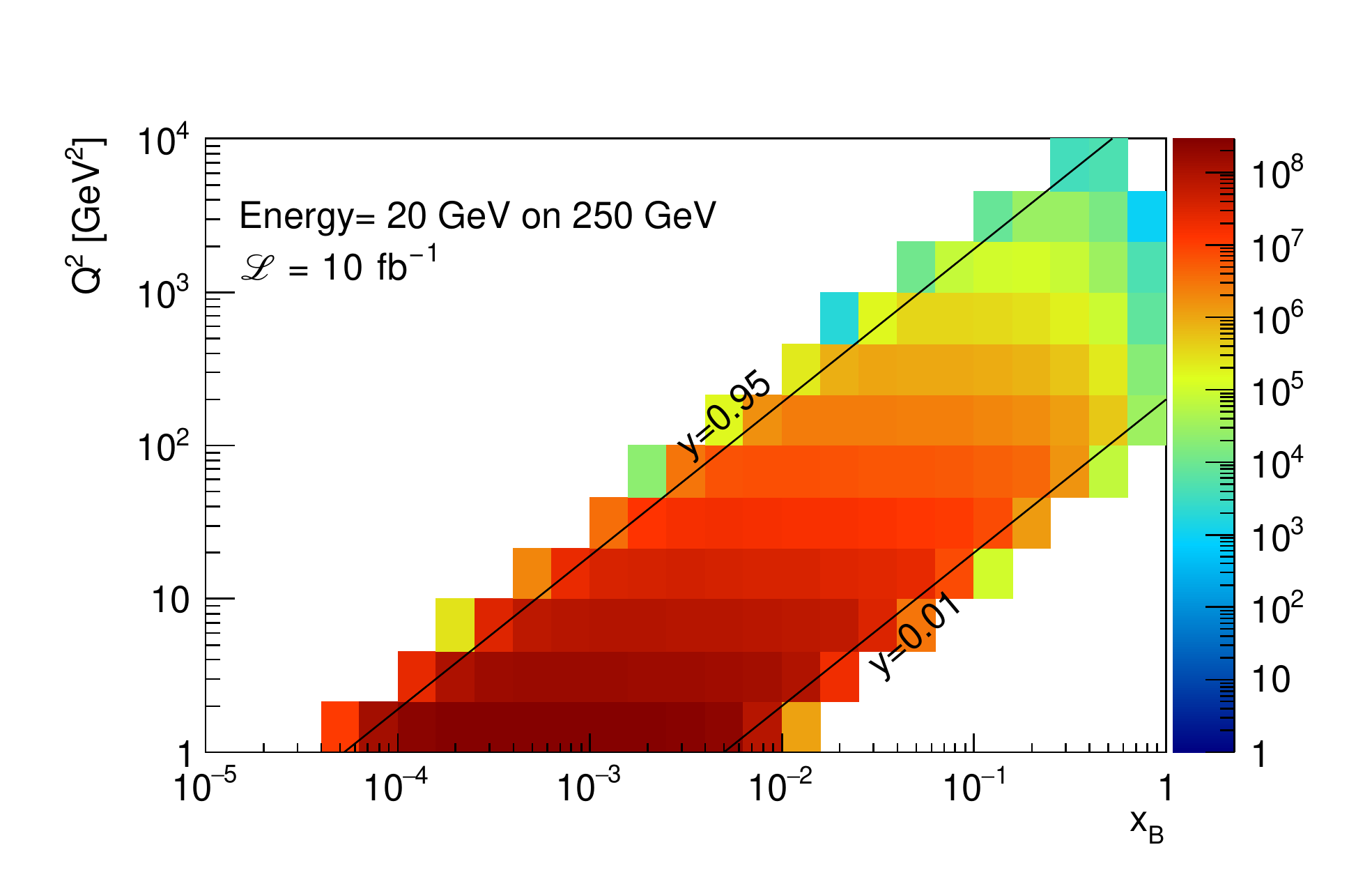,width=1.0\textwidth}
  \end{minipage}
  \caption{Expected distribution of SIDIS events in bins of $x_B$ and $Q^2$ for two
           electron-proton beam energy combinations 5 GeV on 100 GeV (left) and 20 GeV on 250 GeV 
           (right). The two lines indicate the limits on the $x - Q^2$ plane requiring $0.01 < y < 0.95$. 
           All particles are required to be between $-4$ and $4$ in rapidity.} 
  \label{fig:2D}
\end{figure*}

\section{SIDIS Cross Section at NLO}\label{sec:SIDIS}
%
The cross section for the production of a final-state hadron $H$ in deep-inelastic 
electron-nucleon scattering, $e\,N\rightarrow\,e'H\,X$, in the current-fragmentation
region can be written in full analogy to the inclusive deep-inelastic (DIS) case, 
but in terms of the semi-inclusive structure functions $F^{H}_{1}$  and $F^{H}_{L}$ 
\cite{Furmanski:1981cw,Graudenz:1994dq}:
\begin{equation}
\label{eq:SIDIS}
\begin{aligned}
\frac{d\sigma^{H}}{dx_{B}\,dy\,dz}=&\frac{2\pi\alpha^{2}}{Q^{2}}\bigg[\frac{(1+(1-y)^{2})}{y} 2F^{H}_{1}(x_{B},z,Q^{2})+\\
                                   &\frac{2(1-y)}{y}F^{H}_{L}(x_{B},z,Q^{2})\bigg]\, ,
\end{aligned}                                  
\end{equation}
where $x_{B}$, the inelasticity, $y$, and the virtuality of the exchanged photon, $Q^2$, are the usual 
DIS variables, defined in terms of the nucleon, the 
photon and the incoming electron four-momenta, $p_N$, $q_{\gamma}$ and $k_e$, respectively,
\begin{equation}
x_{B}=\frac{Q^{2}}{2 p_N\cdot q_{\gamma}}, \,\,\,\,\,\,\,\,\, y=\frac{q_{\gamma}\cdot p_N}{k_e\cdot p_N}, \,\,\,\,\,\,\,\,\, Q^{2}=-q_{\gamma}^{2},
\end{equation}
while $z$ is the analog of $x_{B}$ for the fragmentation process 
\begin{equation}
z=\frac{p_{H}\cdot p_{N}}{p_{N}\cdot q_{\gamma}},
\end{equation}
which at the lowest order in QCD can be interpreted as the fraction of the fragmenting 
parton momentum carried by the final-state hadron with momentum $p_H$. In the collinear, leading 
twist, NLO approximation, factorisation allows to write the structure functions $F^{H}_{1}$ and 
$F^{H}_{L}$ in Eq.~\ref{eq:SIDIS} as convolutions of the quark and gluon distribution functions in 
the nucleon, denoted respectively as $f_q$ and $f_g$, and the FF $D_j^H$ into the final hadron $H$:
\begin{equation}
\label{eq:F1}
\begin{aligned}
2F^{H}_{1}(x,z,Q^{2})=&\sum_{q,\bar{q}}e_{q}^{2}\Bigg\{f_q(x,Q^{2})D_{q}^{H}(z,Q^{2})\\
 &+\frac{\alpha_{s}(Q^{2})}{2\pi}\bigg[f_q\otimes C^{1}_{qq}\otimes D_{q}^{H}\\
 &+f_q\otimes C_{gq}^{1}\otimes D_{g}^{H}\\
 &+f_g\otimes C_{qg}^{1}\otimes D_{q}^{H}\bigg](x,z,Q^{2})\Bigg\}\, ,
\end{aligned}                                  
\end{equation}
\begin{equation}
\label{eq:FL}
\begin{aligned}
F^{H}_{L}(x,z,Q^{2})=&\frac{\alpha_{s}(Q^2)}{2\pi}\sum_{q,\bar{q}}e_{q}^{2}\bigg[f
_q\otimes C^{L}_{qq}\otimes D_{q}^{H}\\
 &+f_q\otimes C_{gq}^{L}\otimes D_{g}^{H}\\
 &+f_g\otimes C_{qg}^{L}\otimes D_{q}^{H}\bigg](x,z,Q^{2})\, ,
\end{aligned}                                  
\end{equation}
where $C_{ij}^{1,L}$ are the NLO $\overline{\text{MS}}$ coefficient functions 
\cite{Furmanski:1981cw,Graudenz:1994dq,deFlorian:1997zj}. Fragmentation functions 
are sensitive to the flavour structure of the hadron, and thus the choice of  
specific hadrons in the final-state allows to disentangle the contributions of 
the different flavours of quarks.

In recent years, increasingly precise SIDIS measurements have been performed 
\cite{Airapetian:2012ki,Adolph:2016bwc}, which are nicely described by 
PDFs and current FFs at NLO accuracy. Together with the single-inclusive measurements 
in $e^{+}e^{-}$ annihilation \cite{Lees:2013rqd,Leitgab:2013qh} and hadron production in 
proton-proton collisions \cite{Agakishiev:2011dc,Abelev:2014laa}, 
they have allowed the extraction of FFs in global QCD analyses with unprecedented 
precision \cite{deFlorian:2014xna,deFlorian:2017lwf}, updating previous, less comprehensive 
determinations \cite{deFlorian:2007aj}, and bringing FF accuracy closer to that of the better 
determined valence PDFs.

In this paper we restrict ourselves to the case of transverse-momentum--integrated final-state 
hadrons produced in the current-fragmentation region.
The QCD framework to describe transverse-momentum--dependent final-state hadron production 
is known at NLO accuracy \cite{Daleo:2004pn} as well as hadron production in the target 
fragmentation region in terms of fracture functions \cite{Graudenz:1994dq,Daleo:2003xg,Daleo:2003jf}. 
%
\section{Simulated data for SIDIS at an EIC}\label{sec:pseudo}
%

Two pre-conceptual designs for a future high-energy ($\sqrt{s}$: 20 - 100 GeV upgradeable to 
140 GeV) and high-luminosity (10$^{33-34}$ cm$^{-2}$s$^{-2}$) polarised EIC have evolved, using 
existing infrastructure and facilities \cite{Accardi:2012qut}. One proposes to add
an electron storage ring to the existing Relativistic Heavy-Ion Collider (RHIC) complex at 
Brookhaven National Laboratory (BNL) to enable electron-ion collisions. The other pre-conceptual
design proposes a new electron and ion collider ring at Jefferson Laboratory (JLab), 
utilising the 12 GeV upgraded CEBAF facility as the electron injector.

To span most of the kinematic coverage of an EIC, the studies are performed for lepton beam 
energies of 5~GeV and 20~GeV in combination with proton beam energies of 100~GeV and 250~GeV, 
respectively, using the Monte Carlo generator PYTHIA-6~\cite{Sjo01,Sjo08} to simulate DIS events. 
The presented results are based on data with a statistical uncertainty corresponding to an 
integrated luminosity of 10~fb$^{-1}$. 
We consider only events with $Q^2>1$~GeV$^2$, a squared invariant mass 
of the photon-nucleon system $W^2>10$~GeV$^2$, and an inelasticity $0.01<y<0.95$. 
The kinematic range covered in $Q^2$ and $x_B$ is shown in Fig.~\ref{fig:2D} for two c.m.s.
energies. At the highest c.m.s. energy four decades in $Q^2$ are spanned, 
while $x_B$ reaches from $10^{-4}$ to 1.0.
At fixed $Q^2$, higher c.m.s. energies allow to access the lower region in $x_B$, 
while lower c.m.s. energies can give complementary information at higher $x_B$.

\begin{table*}
\footnotesize
\begin{center}
\begin{tabular}{|c c c c |}
\hline
\multicolumn{1}{|c}{rapidity} & \multicolumn{1}{c}{\;\;\;\;pion momentum [GeV]\;\;\;\;} & \multicolumn{1}{c}{\;\;\;\;kaon momentum [GeV]\;\;\;\;} \
& \multicolumn{1}{c|}{\;\;\;\;proton momentum [GeV]\;\;\;\;}\\\hline
$-3.5<\text{rapidity}<-1.0$ (RICH) & $0.5<p_{H}<5.0$ & $1.6<p_{H}<5.0$ & $3.0<p_{H}<8.0$ \\
$-1.5<\text{rapidity}<-1.0$ ($dE/dx$) & $0.2<p_{H}<0.6$ & $0.2<p_{H}<0.6$ & $0.2<p_{H}<1.0$ \\
& & & \\
$-1.0<\text{rapidity}<1.0$ (DIRC and $dE/dx$) & $0.2<p_{H}<4.0$ & $0.2<p_{H}<0.7$ & $0.2<p_{H}<1.1$ \\
                                              &                 & $0.8<p_{H}<4.0$ & $1.5<p_{H}<4.0$ \\
& & & \\
$1.0<\text{rapidity}<3.5$ (RICH) & $0.5<p_{H}<50.0$ & $1.6<p_{H}<50.0$ & $3.0<p_{H}<50.0$ \\
$1.0<\text{rapidity}<1.5$ ($dE/dx$) & $0.2<p_{H}<0.6$ & $0.2<p_{H}<0.6$ & $0.2<p_{H}<1.0$ \\ \hline
\end{tabular}
\end{center}
\caption{Range in hadron (pion, kaon and proton) momentum ($p_{H}$) covered in the various 
rapidity regions by different particle-identification detectors.}
\label{tab:pid}
\end{table*}

For SIDIS measurements, it is critical to identify different hadrons with high efficiency 
and high purity. To cover the widest range in $x_{B}, Q^2, z,$ and the hadron transverse momentum with respect to the virtual photon 
$p_T^H$, it is crucial to integrate  
particle-identification detectors into the EIC detector over a wide range in rapidity. 
We follow in this paper the EIC convention that positive rapidity corresponds to the 
proton-beam direction.
Detailed design studies for a general purpose EIC detector provided the following results 
important for this study.
The magnetic field of the detector is of critical importance for the lowest detectable hadron 
momentum $p_H$ and the achievable momentum resolution especially at large rapidities 
($\eta \sim |3|$). Particle momenta are limited to a minimal value of $0.5$~GeV imposed by 
the presence of a $3$~T magnet for momentum reconstruction.

For this study, we assume particle identification detectors spanning a rapidity range 
$-3.5<\eta<3.5$. We consider to identify pions, kaons and protons 
at low hadron momentum $p_H$ by means of the measurement of energy loss per path length ($dE/dx$) 
and for medium to high hadron momentum $p_H$ by Cherenkov radiation in a Ring Imaging Cherenkov 
(RICH) detector in the backward ($-3.5<\eta<-1$) and forward ($1<\eta<3.5$) 
rapidity regions, while at mid rapidity ($-1<\eta<1$) energy loss in the gas of a time 
projection chamber (TPC) in combination with a Detector of Internally Reflected Cherenkov (DIRC) 
light are considered.
The restrictions on the range of detectable hadron momentum associated with particle 
identification capabilities are specified in table~\ref{tab:pid}.

\begin{figure*}[t]\centering
\begin{minipage}{0.49\textwidth}\centering
\epsfig{file=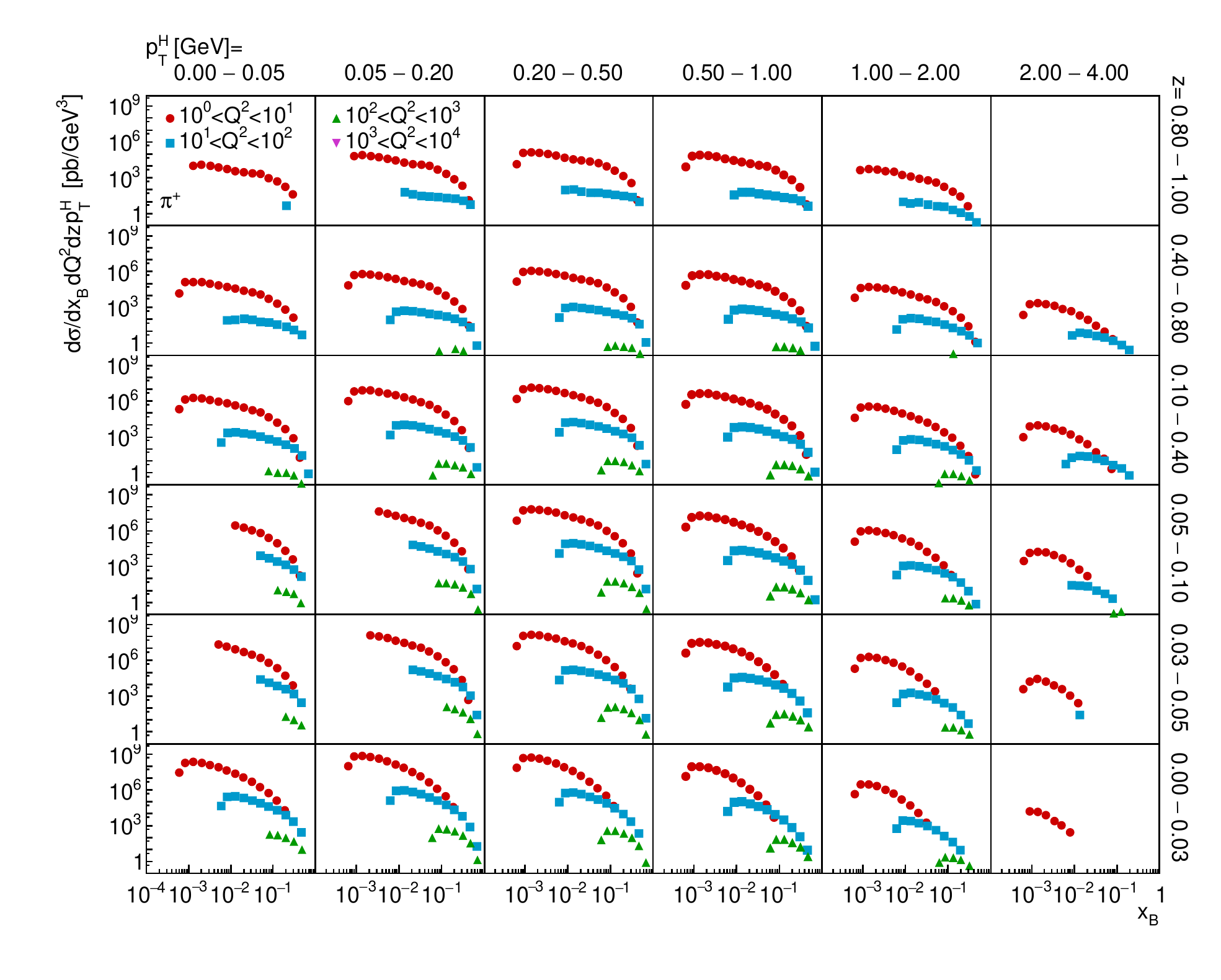,width=1.0\textwidth}
\end{minipage}
\begin{minipage}{0.49\textwidth}\centering
\epsfig{file=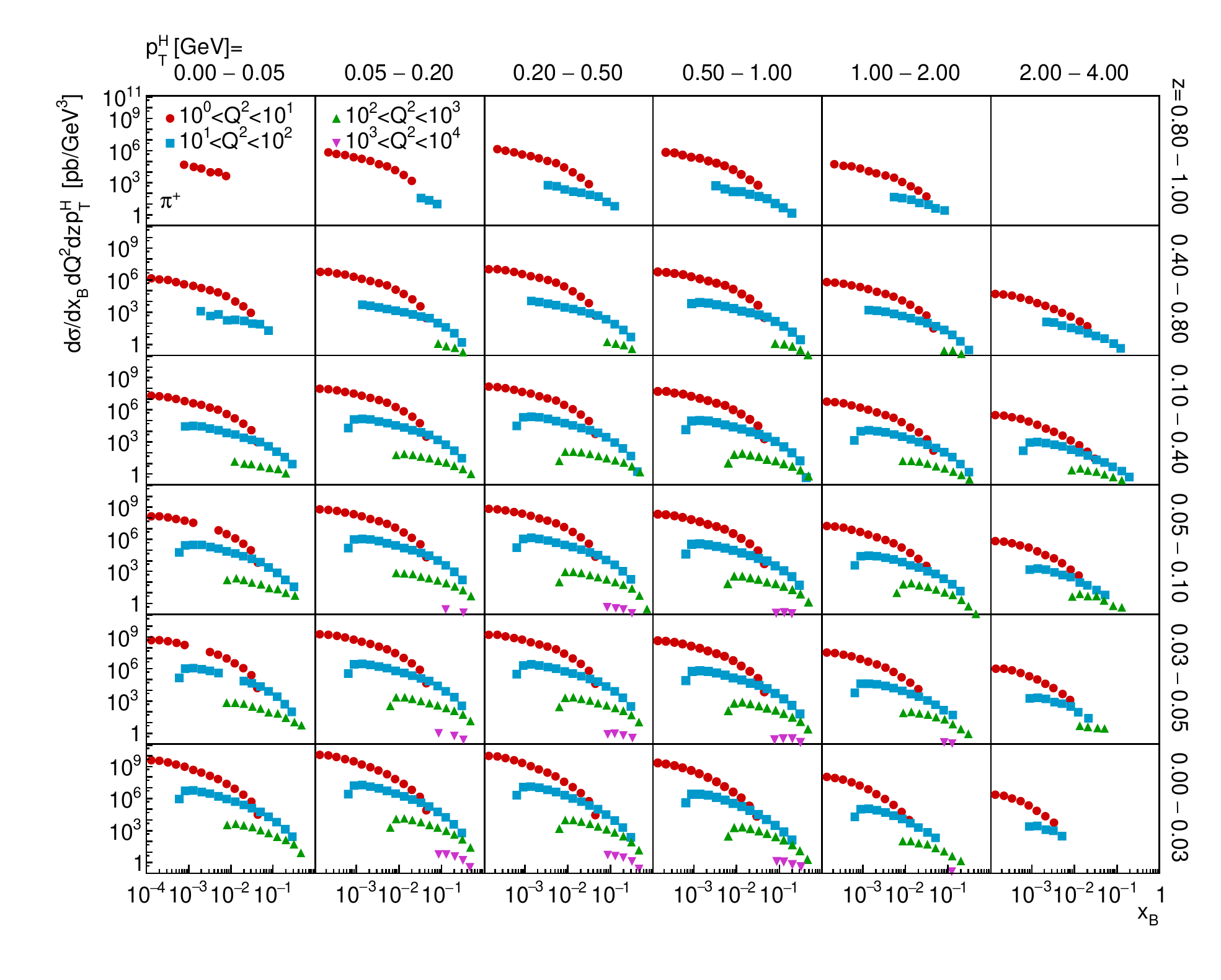,width=1.0\textwidth}
\end{minipage}
\caption{Differential cross section as a function of $x_B$ for bins in $Q^2$, $z$, 
and $p_{T}^H$ for two center-of-mass energies 45 GeV (left) and 140 GeV (right).}
\label{fig:4D}
\end{figure*}

The cross section differential in $x_B$, $Q^2$, $z$, and $p_{T}^H$ for two c.m.s. energies $\sqrt{s}$ = 45 Gev 
and 140 GeV accounting for the above described detector performance is presented in 
Fig.~\ref{fig:4D}. In this figure, the differential cross section is shown for positively
charged pions as a function of $x_B$ for different ranges in $Q^2$, $z$, and $p_{T}^H$. 
Note that a finer binning in $Q^2$ is possible, but for clarity only a sub-division per decade is presented here. 
As already discussed, different beam energies allow to probe complementary regions in 
$x_B$ and $Q^2$ independent of $z$ and $p_{T}^H$. 
Measurements of SIDIS at an EIC will give access to extremely low $p_{T}^H$ and $z$.

\begin{figure}
\epsfig{file=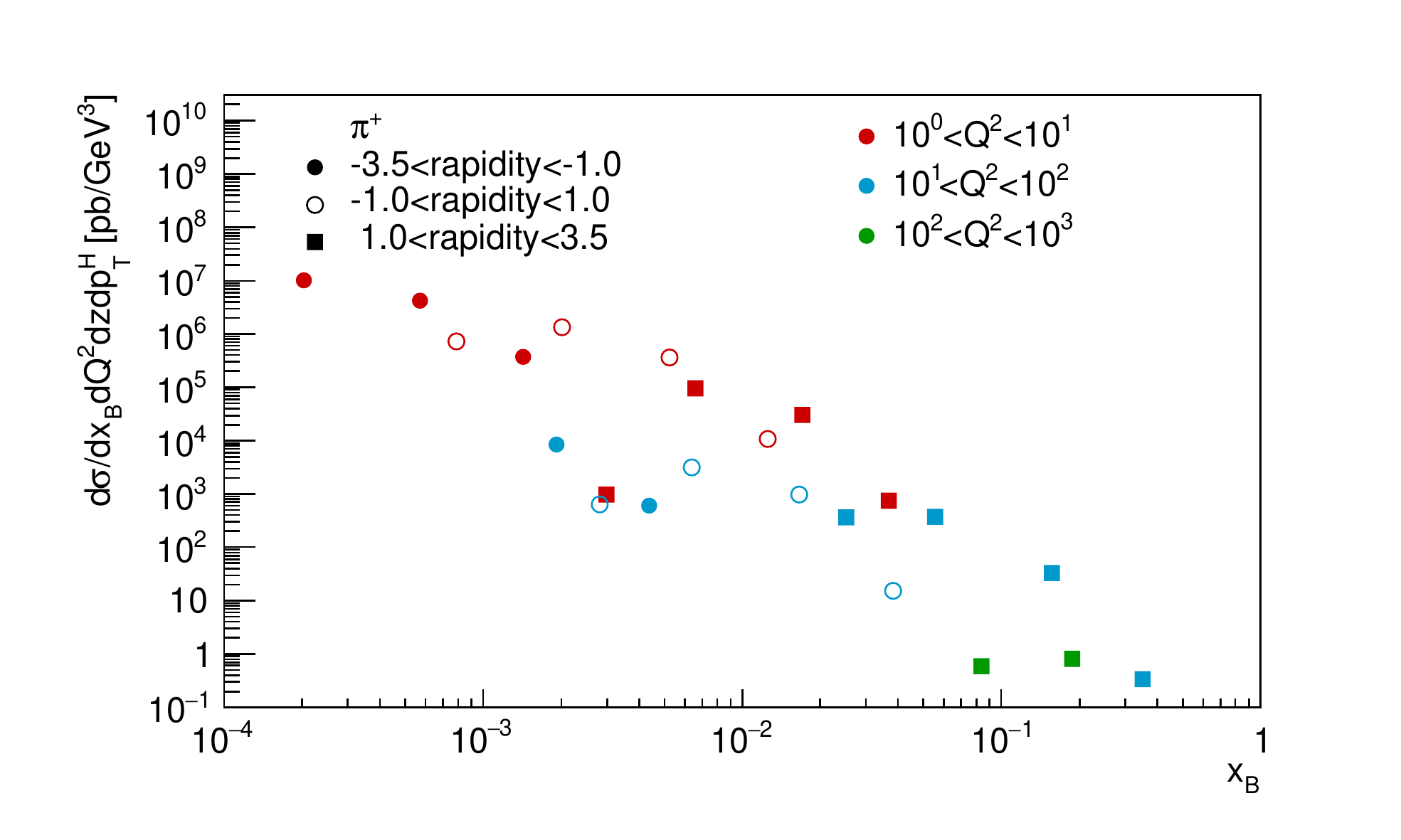,width=0.49\textwidth}
\caption{The simulated differential cross section for three bins in $Q^2$ (indicated 
by the different colors) as a function of $x_B$ for $0.4<z<0.8$ and $0.2<p_{T}^H<0.5$ and all
PID detector requirements (see Table \ref{tab:pid} applied, separated in three rapidity ranges, 
backward rapidity -3.5$<\eta<$-1 (filled circles), mid rapidity -1$<\eta<$1 (open circles), and forward 
rapidity 1$<\eta<$3.5 (filled squares).} 
\label{fig:eta_cov}
\end{figure}

The advantage of particle detection and identification over a large range in rapidity is 
illustrated in Fig.~\ref{fig:eta_cov}, where the four-differential cross section for pion 
production is shown for the three rapidity ranges $-3.5<\eta<-1.0$, $-1.0<\eta<1.0$, and 
$1.0<\eta<3.5$, and for different ranges in $Q^2$, at $\sqrt{s}$=140 GeV.
The pion fractional energy and transverse momentum are limited for this figure to $0.4<z<0.8$ 
and $0.2<p_{T}^H<0.5$. All particle-identification cuts as listed in Table \ref{tab:pid} are applied.
As can be seen, the lower $Q^2$ region is accessed at backward rapidity, while higher $Q^2$ values 
are reached at forward rapidity.
At fixed $Q^2$, lower values of $x_B$ are covered at backward rapidity, while the higher-$x_B$ 
region is probed at forward rapidity. Hence, providing particle identification at backward, 
mid, and forward rapidity is important to cover the widest range in $x_B$ and $Q^2$ possible.

\begin{figure*}\centering
\begin{minipage}{0.3\textwidth}\centering
\epsfig{file=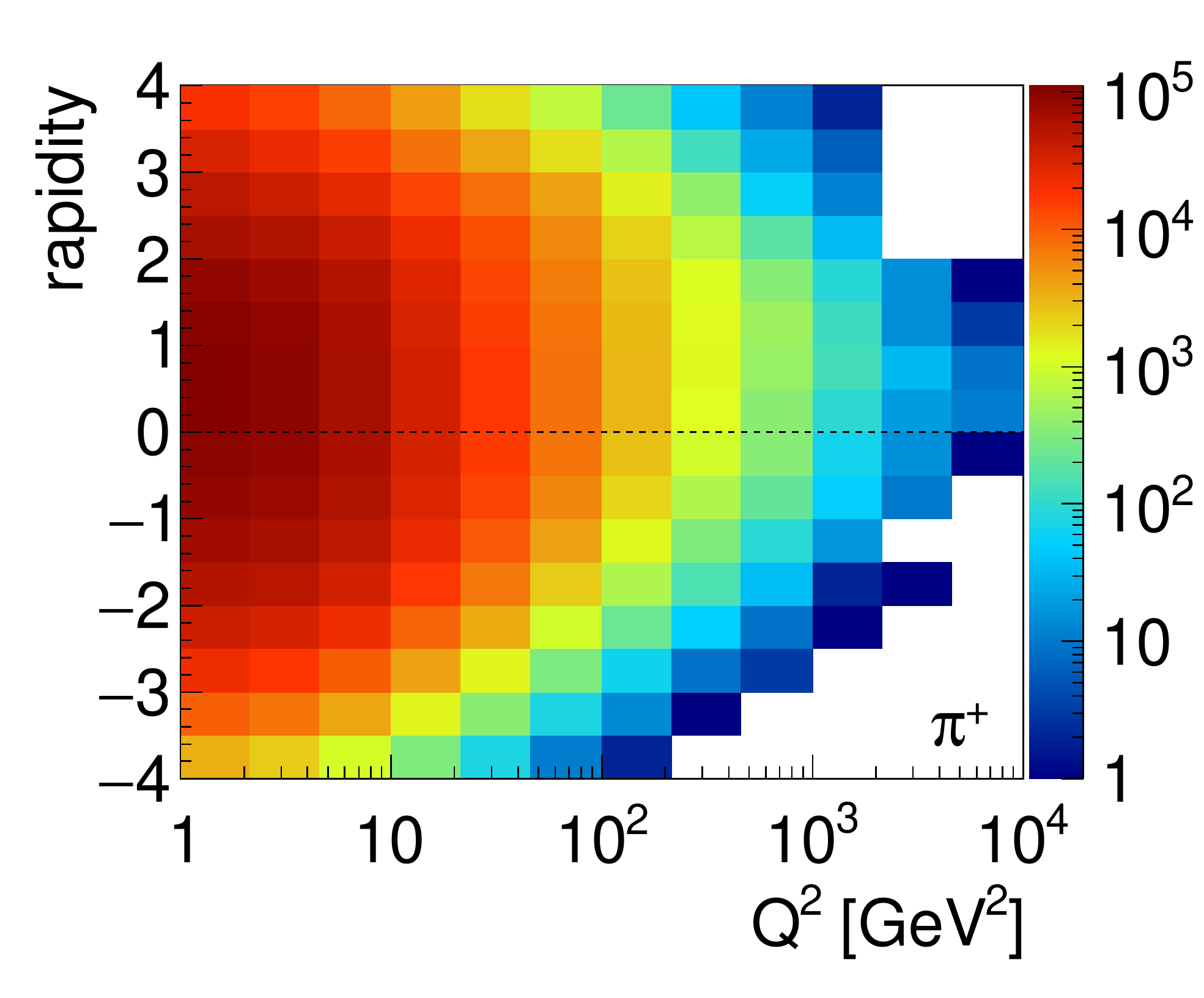,width=1.0\textwidth}
\epsfig{file=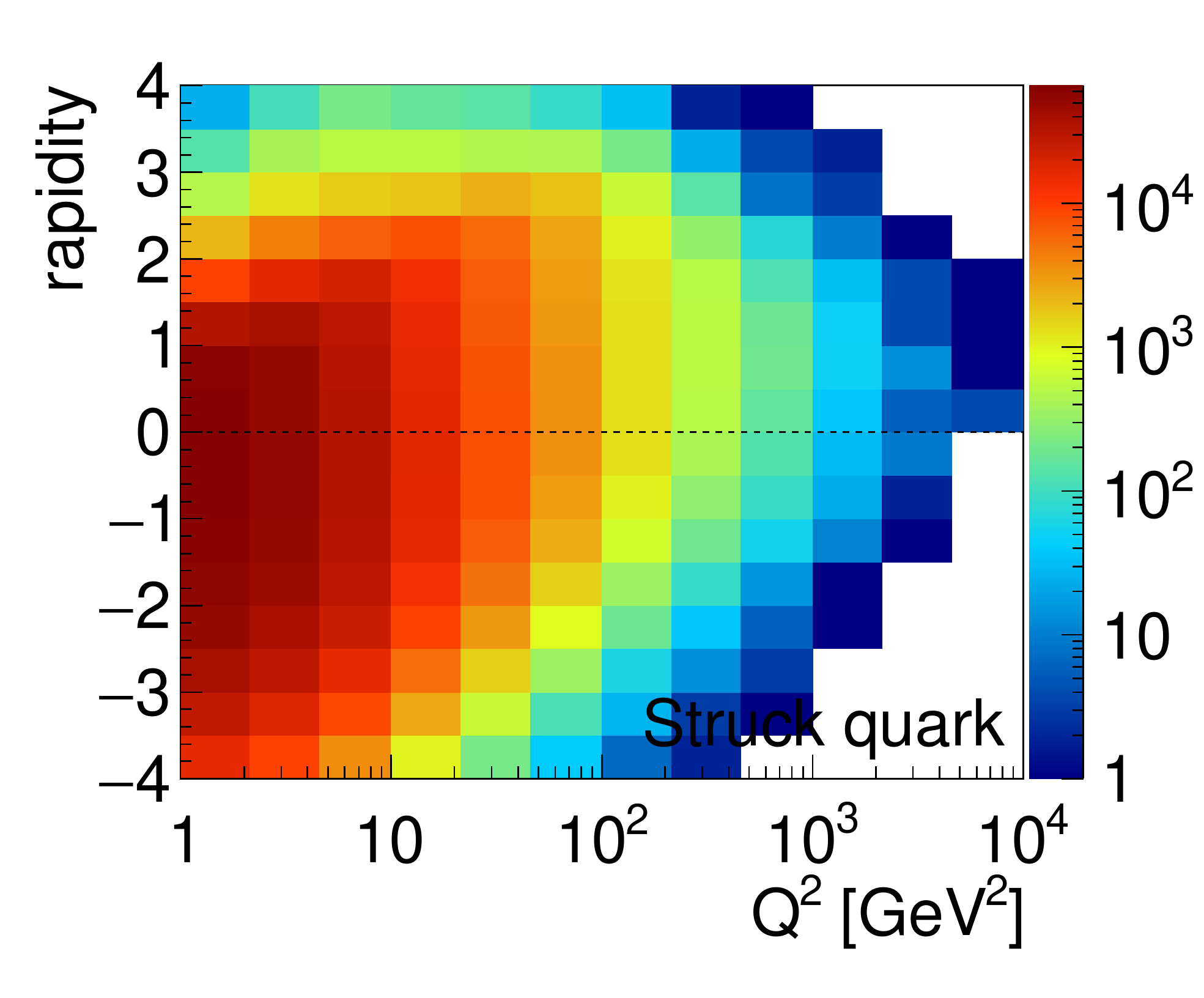,width=1.0\textwidth}
\end{minipage}
\begin{minipage}{0.3\textwidth}\centering
\epsfig{file=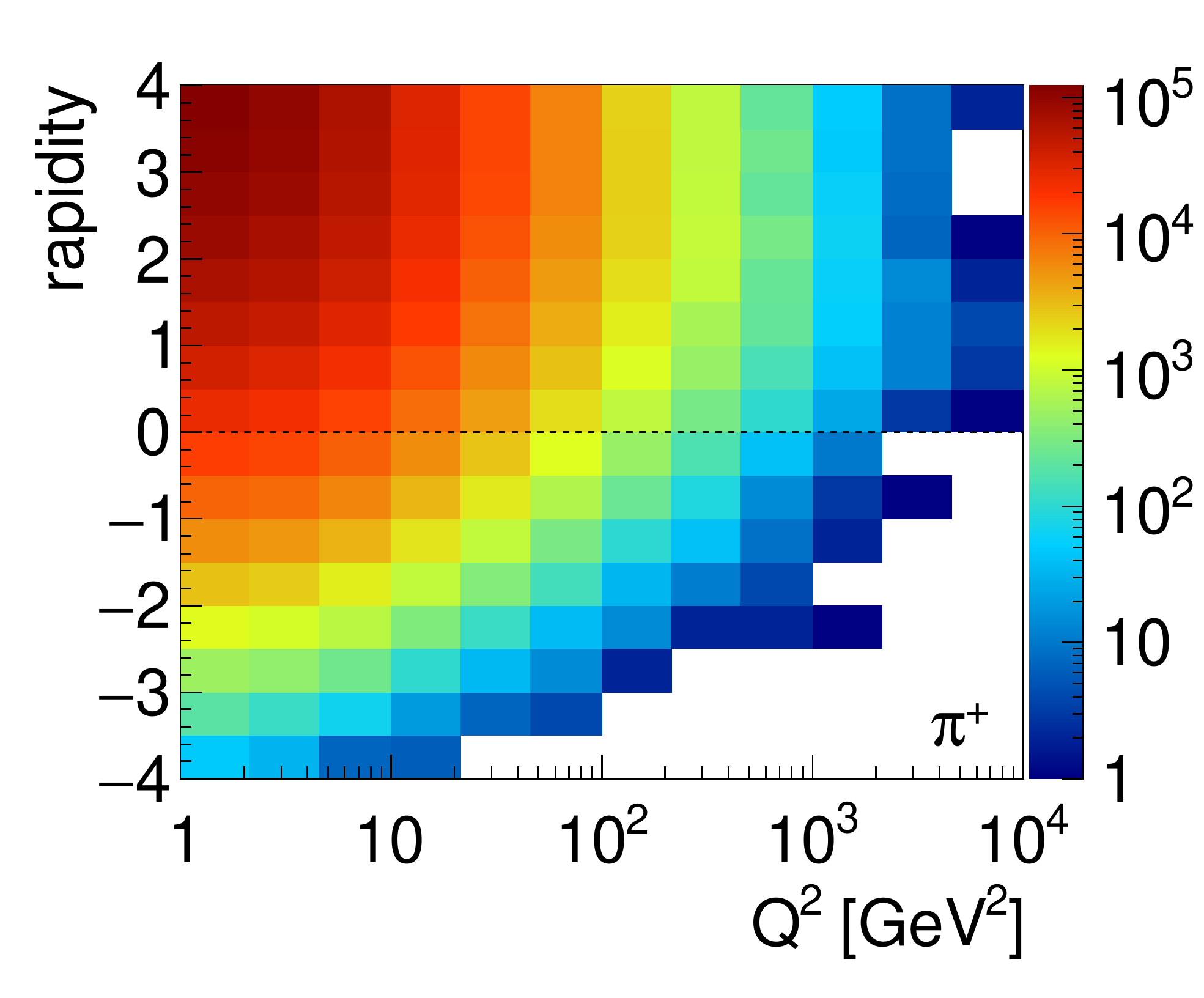,width=1.0\textwidth}
\epsfig{file=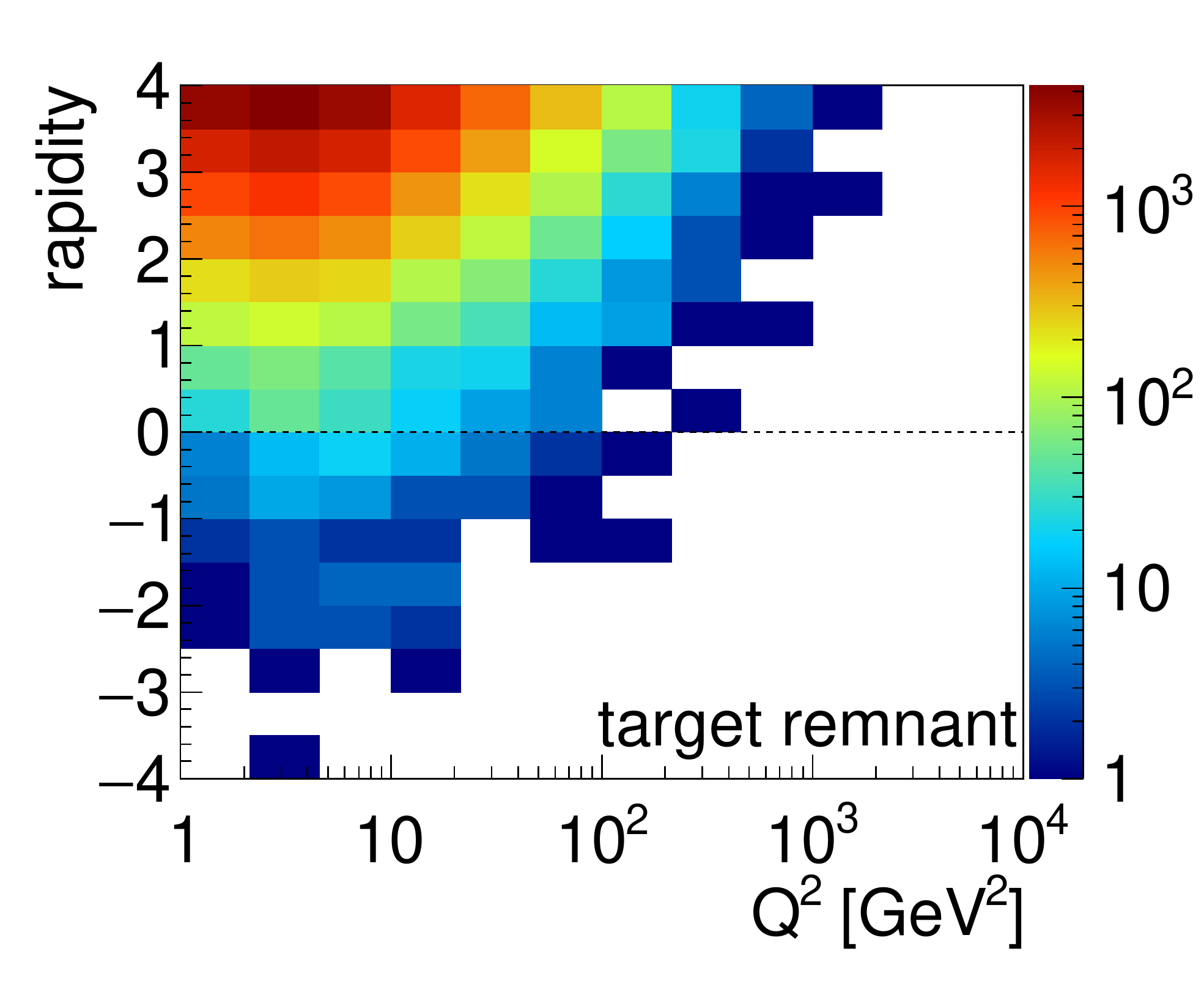,width=1.0\textwidth}
\end{minipage}
\caption{Kinematic range in $Q^2$ and rapidity at $\sqrt{s}$ = 140 GeV for pions originating 
from a struck quark (top, left), the target remnant (top, right), for the struck quark (bottom, left) 
and the target remnant (bottom right) for the DIS subprocess $\gamma^{*}q\rightarrow q$ in PYTHIA-6.}
\label{fig:q-pi}
\end{figure*}
\begin{figure*}\centering
\begin{minipage}{0.3\textwidth}\centering
\epsfig{file=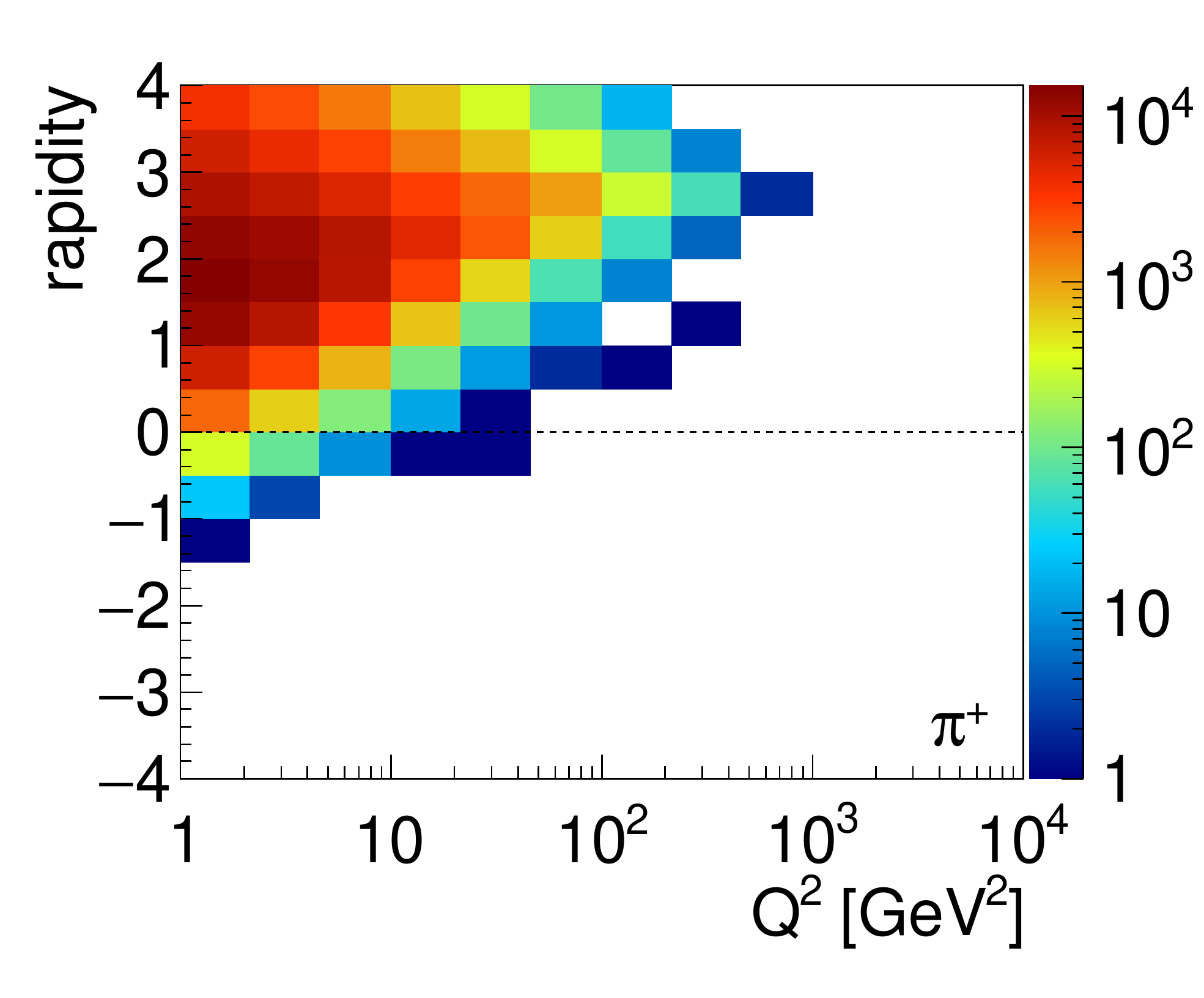,width=1.0\textwidth}
\end{minipage}
\begin{minipage}{0.3\textwidth}\centering
\epsfig{file=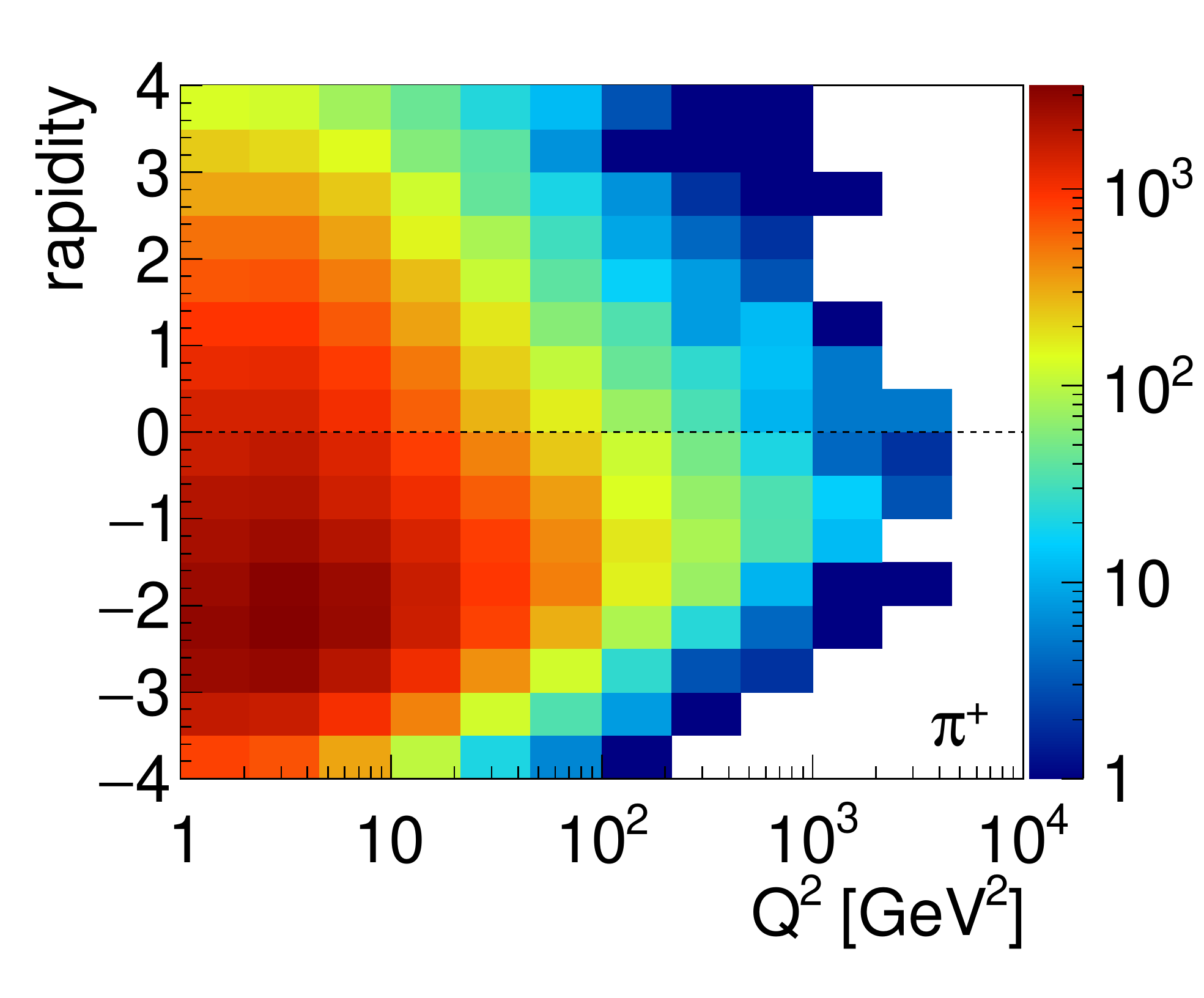,width=1.0\textwidth}
\end{minipage}
\caption{Kinematic range in $Q^2$ and rapidity for pions originating from a struck quark 
for the DIS subprocess $\gamma^{*}q\rightarrow q$ in PYTHIA-6 with 
$200$~GeV$^2<W^2<300$~GeV$^2$ (left) and $16000$~GeV$^2<W^2<18000$~GeV$^2$ (right) for 
$\sqrt{s}$ = 140 GeV.}  
\label{fig:q-pi-W}
\end{figure*}
\begin{figure*}\centering
\epsfig{file=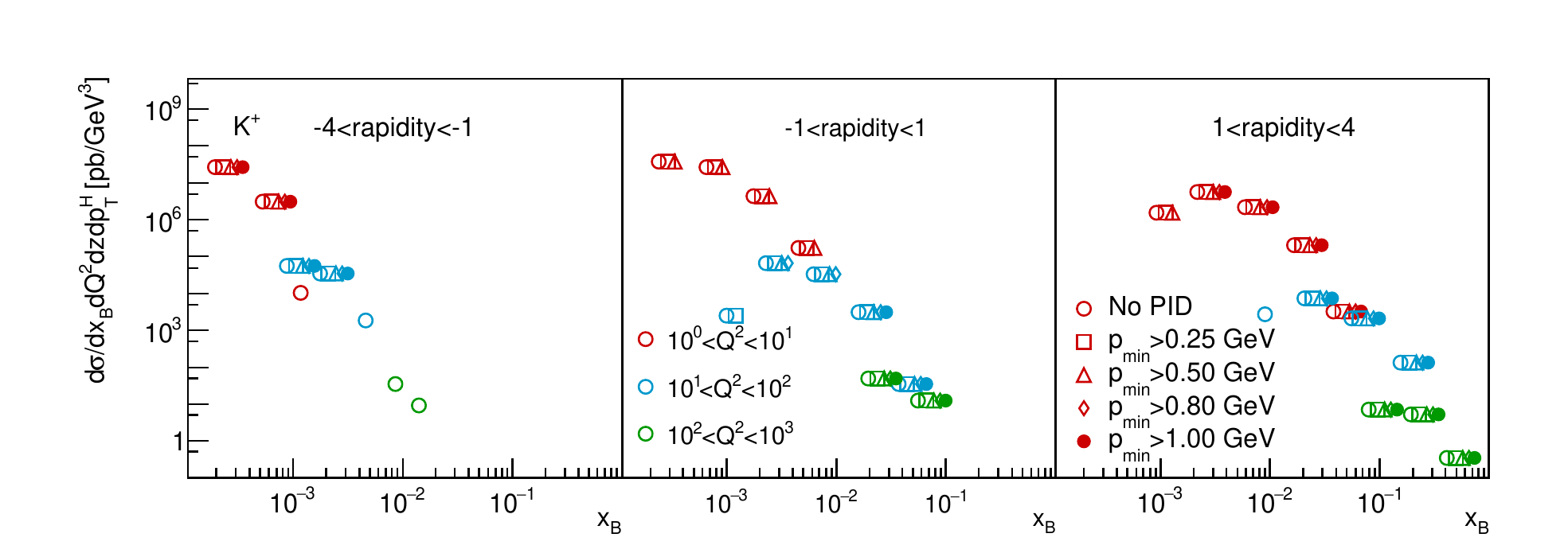,width=1.0\textwidth}
\caption{Differential cross section for kaon production in three regions in rapidity, without 
particle-identification and minimum-momentum requirement (open circles), compared to the results 
with particle-identification and for different detectable lowest particle momenta.}
\label{fig:ploss}
\end{figure*}

An important point about probing various regions in rapidity is the enhanced lever-arm to separate 
current and target fragmentation. This is illustrated in Fig.~\ref{fig:q-pi}.
Here, the upper panels show the distribution of pions originating from a struck quark (left) 
and from the target remnant (right) in the $Q^2$-rapidity plane for the DIS subprocess 
$\gamma^{*}q\rightarrow q$ in PYTHIA-6.
The bottom panels show the distributions of the struck quark (left) 
and the target-remnant (right) from which the pion originates\footnote{The struck quark is 
selected using internal PYTHI-6 information by cutting on the status code KS equal to 11 or 12, 
and the parent particle with KS=21. The target remnant was selected requiring either a quark or 
a di-quark through KS=11 or 12 and the nucleon as parent particle.}. 
While one has to be very careful with the interpretation of the classification of hadrons 
and their origin in Monte Carlo generators, this plot illustrates clearly that
there exists a correlation between the direction of a hadron and its origin. 
As expected, target remnants are populating regions in rapidity that are much more forward 
than what is correlated with the struck quark, and its associated pions follow the same trend.
While the correlation is not 100\% and in reality many more sub-processes than the one 
exemplified here contribute, the figure illustrates that by covering different regions in 
rapidity, one can obtain an improved separation of current and target fragmentation.
Note that this correlations reveal also a clear $W^2$ dependence, as shown 
for two regions in $W^2$ in Fig.~\ref{fig:q-pi-W}.

While particle-identification detectors will most likely not allow for a full coverage 
in acceptance, they should be chosen to provide a minimal loss in kinematic coverage. 
Similarly, the choice of the magnet strength is a compromise between the loss of 
low-momentum, i.e., low-$z$ hadrons, the fraction 
of which increases with increasing magnetic field, and the degradation in momentum resolution 
at high momenta, which is inversely proportional to the strength of the magnetic field.
The kinematic regions, where particles are lost due to particle-identification requirements and 
the presence of a magnetic field are shown in Fig.~\ref{fig:ploss} for positively charged kaons 
for $\sqrt{s}$ = 140 GeV. The open circles correspond to the cross section not requiring a lower 
momentum cut due to the magnetic field and no restriction due to particle-identification in the 
rapidity range between -4 and 4. 
All other symbols represent the situation requiring particle identification, 
as detailed in table~\ref{tab:pid}, and a different lower momentum cut-offs.
As seen from the figure, data at higher $x_B$ values are lost at backward rapidity,
because of the particle-identification requirements. However, the same kinematic region is 
accessible at mid rapidity, if the minimal momentum cut can be below 0.80~GeV.
The complementarity offered by the various rapidity ranges, provided they are equipped with 
the appropriate detector components, is clearly illustrated in this figure. 
For the lower center-of-mass energy the same conclusions hold for pions, kaons and protons.

\begin{figure*}\centering
\epsfig{file=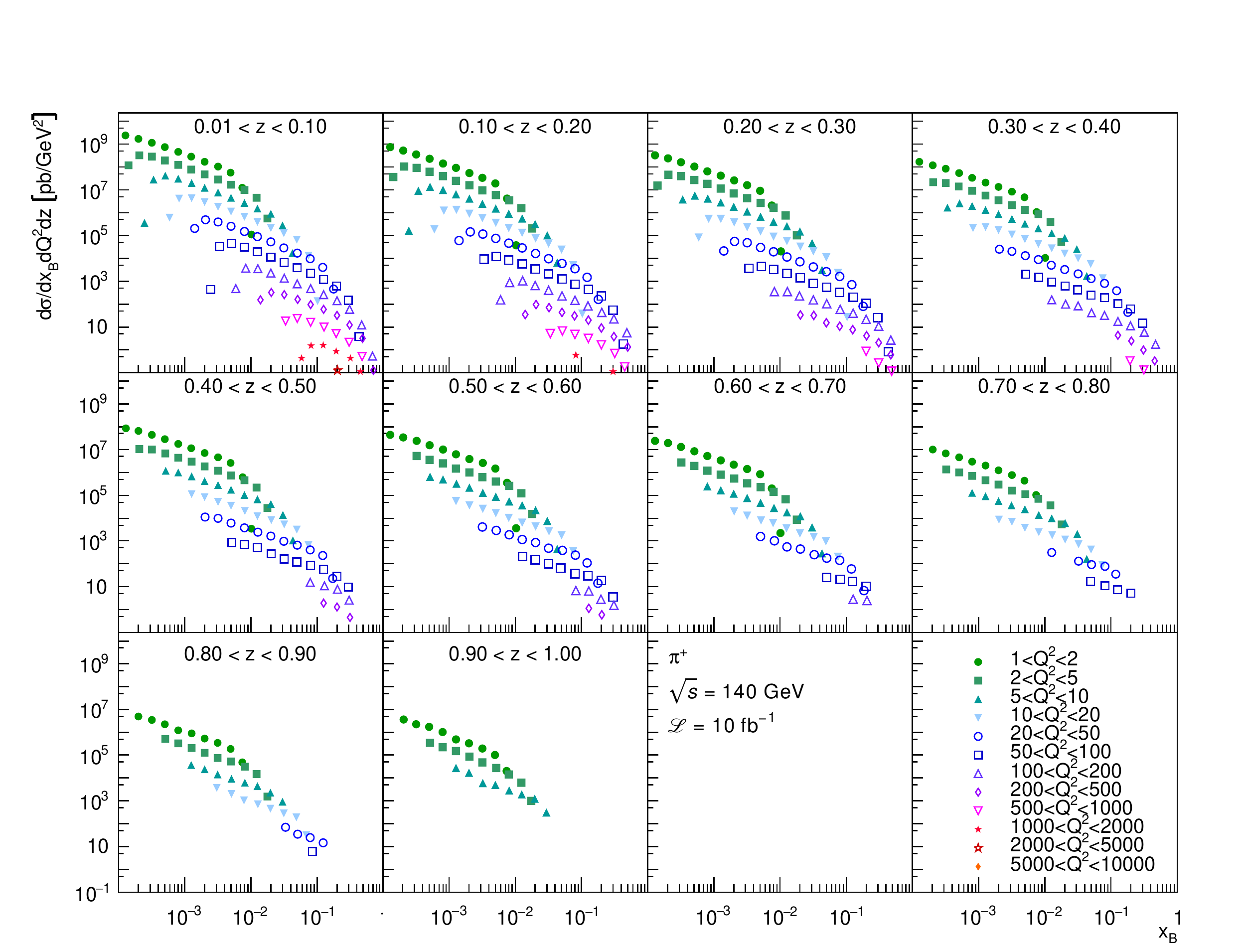,width=1.0\textwidth}
\caption{Differential cross section for pion production at $\sqrt{s}$ = 140 GeV as a function of 
$x_B$ for bins in $Q^2$ and $z$ measurable at an EIC.} 
\label{fig:3Dxsec}
\end{figure*}

In the following all impact studies for PDFs and FFs are performed based on simulated 
data that satisfy DIS and particle-identification requirements for hadrons from 
table~\ref{tab:pid}, with a lower momentum cut off of 0.5~GeV. 
The corresponding cross section as function of $x_B$ binned in $Q^2$ and $z$ 
unfolded for detector effects is illustrated in Fig.~\ref{fig:3Dxsec} 
for pions at a c.m.s. energy of 140~GeV. The uncertainties correspond to a integrated
luminosity of 10 fb$^{-1}$. 

Besides the statistical uncertainties one would need to also consider the systematic uncertainties. 
They consist of an overall systematic uncertainty of 1.4\% on the luminosity determination and 
a bin-by-bin systematic uncertainty to account for the challenges to identify hadrons 
over a wide kinematic range and any other detector effects, which cannot be fully unfolded. A current conservative 
estimate of the bin-by-bin systematic uncertainty is 3.5\%. It should be added in quadrature
to the statistical one. As it is difficult without a full detector design to estimate this 
bin-to-bin uncertainty reliably, we decided to not consider it in our study. 

%
\section{Bayesian and Hessian toolbox}\label{sec:status}
\subsection{PDF and FF reweighting with SIDIS data}
%
One of the key ingredients in the strategy pursued in the present analysis is the use of 
reweighting methods to a set of PDFs or FFs, as a means 
to incorporate additional information from new data into an existing set, 
without the need to perform a new global fit \cite{Ball:2010gb,Ball:2011gg}. 
Successful demonstrations of the method have been performed in different applications, 
and more specifically, its usefulness in constraining PDFs with
actual SIDIS data has already been shown in \cite{Borsa:2017vwy}. 
Here, we briefly recall the main features that are needed for our analysis below.

The method was originally developed based on Bayesian inference, and relies on the beforehand 
generation of a large ensemble of PDF or FF sets $f_{i}^{(k)}$, by fitting replicas of data 
obtained by smearing available experimental data according to their experimental and systematic 
uncertainties and correlations. 
Here, $i$ is indexing the parton flavour and $k$ the numbers the replica. The such obtained set 
of PDF or FF replicas forms a precise representation of the underlying probability distribution 
for the PDFs or FFs. Any observable ${\cal O}$ depending on PDFs and FFs can be evaluated 
by averaging the results for the individual replicas:
\begin{equation}
\label{eq:mean}
\langle{\cal O} \rangle = \frac{1}{N} \sum_{k=1}^{N}{\cal O}[f_i^{(k)}]\,,
\end{equation}
with $N$ the number of replicas, and the corresponding variance defined as
\begin{equation}
\label{eq:sigma}
\Delta {\cal O} = \sqrt{ \frac{1}{N-1} \sum_{k=1}^N \left({\cal O} [f_i^{(k)}] - \langle{\cal O} \rangle \right)^2  } 
\end{equation}

Using Bayesian inference, it is possible to assess the effect of a new, independent data set 
by updating the probability distribution through the assignment of a new weight 
$w_k\neq 1$ for each replica. 
This weight measures the agreement of replica $k$ with the new data. 
The weighted estimate for any observable then becomes
\begin{equation}
\label{eq:newmean}
\langle {\cal O} \rangle_{\text{new}} = \frac{1}{N} \sum_{k=1}^{N} w_k\,{\cal O}[f_i^{(k)}]\,,
\end{equation}

Clearly, replicas with very small weights become irrelevant in the calculation of any 
observable, thus reducing the spread of the modified probability distribution compared 
to the original one. As long as the new data set is not too restrictive, and the number 
of replicas with non-negligible values of $w_{k}$ is large enough, reweighted PDFs or FFs 
will form an accurate representation of the original probability distribution.

The reweighting strategy can also be implemented within the Hessian approach for uncertainties
in global PDF or FF extractions \cite{Paukkunen:2014zia}. In this case, the large ensemble 
of replicas needed can be constructed as a gaussian smearing of the Hessian eigenvector sets:  
\begin{equation}
\label{eq:hessian_replicas}
f_{k}\equiv f_{S_{0}}+\sum_{i}^{N_{eig}}\bigg(\frac{f_{S_{i}^{+}}-f_{S_{i}^{-}}}{2}\bigg) R_{ik}\,.
\end{equation}

Here, $f_{S_{0}}$ corresponds to the value of the PDF (FF) obtained with the best fit parameters, 
while $f_{S_{i}^{+}}$ and $f_{S_{i}^{-}}$ are the values of the PDF (FF) evaluated 
for extreme displacements in the direction of the $i-th$ eigenvector. $R_{ik}$ are random 
numbers with a Gaussian distribution centered at zero and with variance one. The weights 
$w_{k}$ for each replica can be calculated in a completely analogous way as in the case of 
Monte-Carlo based replicas, and therefore, the reweighted PDF (FF) can be written as
\begin{equation}
\label{eq:hessian_replicas_rew}
f_{new}\equiv f_{S_{0}}+\sum_{i}^{N_{eig}}\bigg(\frac{f_{S_{i}^{+}}-f_{S_{i}^{-}}}{2}\bigg) \Bigg(\frac{1}{N_{rep}}\sum_{k}^{N_{rep}}w_{k}\,R_{ik}\Bigg)\,.
\end{equation}

In the following, we use an ensemble of 1000 PDF replicas from \cite{Ball:2014uwa} to perform 
the PDF reweighting, while in the case of the FFs a set of $10^{5}$ replicas is generated from 
Eq.~\ref{eq:hessian_replicas}.  We compute the weights by comparing to which extent each replica $k$ 
reproduces the EIC SIDIS pseudodata for charged pions and kaons. The much larger number of starting 
FF replicas is related to the fact that current sets of FFs are typically much less constrained 
than PDFs and the reweighting with very precise data as that expected from EIC leaves 
a comparatively small number of surviving replicas. The SIDIS cross sections are computed 
at NLO accuracy by 
convoluting each replica with a variant of the DSS FFs for pions and kaons 
\cite{deFlorian:2014xna,deFlorian:2017lwf}, but upgraded so that they use the 
NNPDF3.0 set of PDFs and the corresponding $\alpha_s$ as input for consistency \cite{Borsa:2017vwy}. 
 
Notice that there is a subtlety regarding the inclusion of SIDIS data in the reweighting 
procedure, since in addition to the experimental uncertainties of the pseudodata, 
there are also uncertainties associated to the FFs when reweighting PDFs, or conversely 
to PDFs when reweighting FFs, which are used in the calculation of the observable. These of course 
have to be taken into account when computing the weights. For the FF reweighting, the 
uncertainties associated to the PDFs are added in quadrature to the experimental uncertainties, 
and for the PDF reweighting, the FF uncertainties are included in a similar way. 
The latter case is more involved, since the FF uncertainty estimates already include those 
of the PDFs used in the original FF extraction, producing a mild double counting that needs 
to be accounted for.   
This issue was addressed in \cite{Borsa:2017vwy}, where a criterion on how to include the FF 
uncertainty consistently was proposed. In the following we adopt the same procedure. 
%
\subsection{Correlations with Monte Carlo replicas}\label{sec:correlation}
%
Another major advantage of Monte Carlo replicas and Hessian eigenvector sets lies in the 
possibility to use them in order to scan the regions of phase space where the measurements 
for some observable can potentially constrain the non-perturbative distributions (PDFs and FFs). 
This can be achieved through the calculation of the correlation coefficients between that 
observable and the PDF (FF) for a given flavour. The calculation of correlations both in the 
Hessian and the Monte Carlo formalisms has been discussed in detail in the literature 
\cite{Bertone:2018ecm,Nadolsky:2008zw,Ball:2008by,Guffanti:2010yu,Wang:2018heo}. 

In the case of a set of replicas for PDFs based on the Monte Carlo method, the correlation 
coefficient $\rho\,[f_i ,{\cal O}]$ between a PDF for a given flavour $i$ and an observable 
${\cal O}$ (i.e. the cross section for a given process) can be defined as \cite{Guffanti:2010yu} 
\begin{equation}
\label{eq:pdf_correlation}
\rho\,[f_i,{\cal O}]=\frac{\langle {\cal O}\cdot f_i\rangle -\langle {\cal O}\rangle \langle f_i\rangle }{\Delta{\cal O} \Delta{f_i}}\, ,
\end{equation}
where the mean values are calculated over the ensemble of replicas as in Eq.~\ref{eq:mean}, 
while the standard deviations for the observable and parton density are given by Eq.~\ref{eq:sigma}. 
Values for $|\rho|$ close to unity indicate that the observable and the PDF are highly 
correlated and therefore, including data of that type with  competitive experimental 
uncertainties could in principle further constrain the PDF. Values close to zero are obtained 
for uncorrelated observables, which would never be able to improve the PDF determination,
irrespective of how precise that data are. For simplicity, we omit the dependencies on 
$x_{B}$, $Q^{2}$ and $z$, however, the correlation coefficients are defined for the kinematics 
of each individual point of the pseudodata, allowing a straightforward comparison 
between the constraining power of different kinematics.
 
It is noted that the correlation coefficients can only give insight into the {\em potential} impact 
that the new data could have on the PDF or FF determination, but they do not take into 
account the experimental uncertainties for the observable, which ultimately determine the 
{\em actual} constraining power. If for a given region of phase space the experimental uncertainties 
are large compared to the uncertainty propagated from the PDFs, it is reasonable to expect 
that these measurements will not constrain the PDFs in this region, regardless of the value of the 
correlation coefficient. 

In order to have a better estimate of the impact of the actual data in a global fit, one can
define a scaled correlation or sensitivity coefficient \cite{Wang:2018heo} as
\begin{equation}
\label{eq:pdf_correlation_esc}
S[f_i,{\cal O}]=\frac{\langle {\cal O}\cdot f_i\rangle -\langle {\cal O}\rangle\langle f_i\rangle}{\xi\,\Delta{\cal O}\Delta{f_i}}\, ,
\end{equation}
where the scaling factor
\begin{equation}
\label{eq:xi}
 \xi \equiv \frac{\delta {\cal O}}{\Delta {{\cal O}}}
\end{equation}
is defined as the ratio of the experimental uncertainties of the measurement $\delta{\cal O}$ 
and the theoretical uncertainty for that same observable propagated from the PDFs $\Delta{\cal O}$. 
The scaled correlation coefficient suppresses those regions of phase space for which the 
experimental uncertainties are large compared to the uncertainty associated to the PDFs, 
while it enhances those regions where the largest impact on the distributions is expected. 
Of course, the scaled coefficients are no longer constrained to vary within $[-1,1]$. 

\subsection{Correlations within the Hessian approach}

While several sets of PDF replicas based on the Monte Carlo method are nowadays available, this 
is not the case for the FFs.  In \cite{Bertone:2018ecm}, Monte Carlo based FFs have been produced, 
however they do not include charge separation, neither do they include SIDIS data. 
On the other hand, extractions like those in \cite{deFlorian:2014xna,deFlorian:2017lwf} 
include flavour separation and SIDIS data, estimate uncertainties using the Hessian 
strategy such the previous method can not be directly applied. Nevertheless, it is still 
possible to quantify the correlations within the Hessian formalism. 
One can define a correlation coefficient analogous to $\rho\,[f_i,{\cal O}]$, 
in terms of Hessian eigen-vector sets following \cite{Wang:2018heo}:
\begin{equation}
\label{eq:ff_correlation}
\rho\,[D_{q}^{H},{\cal O}]=\frac{\vec{\nabla}D_{q}^{H} \cdot \vec{\nabla}{\cal O}}{\Delta D_{q}^{H}\,\Delta {\cal O}},
\end{equation}
where the gradient is taken in the space of Hessian eigenvector FF parameters, and can 
be approximated by this finite-difference
\begin{equation}
\label{eq:hessian_gradient}
\frac{\partial X}{\partial x_{i}}=\frac{1}{2}(X_{i}^{+}-X_{i}^{-}),
\end{equation}
where $X_{i}^{\pm}$ represents the values of $X$ for extreme displacements along the 
direction of the $i-th$ eigenvector, for a given tolerance.
Similarly, the uncertainty for any observable can be estimated as:
\begin{equation}
\label{eq:hessian_error}
\Delta X=|\vec{\nabla}X|=\frac{1}{2}\sqrt{\sum_{i=1}^{N}(X_{i}^{+}-X_{i}^{-})^{2}},
\end{equation}
so that the expression for the correlation in Eq.~\ref{eq:ff_correlation} can be recasted as
\begin{equation}
\label{eq:hessian_corr}
\rho\,[D_{q}^{H},{\cal O}]=\frac{1}{4\,\Delta D_{q}^{H}\,\Delta {\cal O}}\sum_{i=1}^{N}\big[(D_{q}^{H})_{i}^{+}-(D_{q}^{H})_{i}^{-}\big]({\cal O}_{i}^{+}-{\cal O}_{i}^{-}).
\end{equation}
As in the case of the PDF correlations, it is worth noting that the correlations defined 
in Eq.~\ref{eq:hessian_corr} do not account for the experimental uncertainties of the new 
data nor the precision already achieved in the distributions, so it is convenient to define 
a sensitivity coefficient \cite{Wang:2018heo}:
\begin{equation}
\label{eq:hessian_corr_esc}
\begin{aligned}
S[D_{q}^{H},{\cal O}]=&\frac{1}{4\,\xi\,\Delta D_{q}^{H}\,\Delta {\cal O}}\times\\
&\sum_{i=1}^{N}\big[(D_{q}^{H})_{i}^{+}-(D_{q}^{H})_{i}^{-}\big]({\cal O}_{i}^{+}-{\cal O}_{i}^{-})\,,
\end{aligned}
\end{equation}
where again $\xi$ is given by Eq.~\ref{eq:xi}.
%
\section{Results}\label{sec:results}
\subsection{Correlations}\label{subsec:correlation}
%
\begin{figure*}[h]
\epsfig{figure=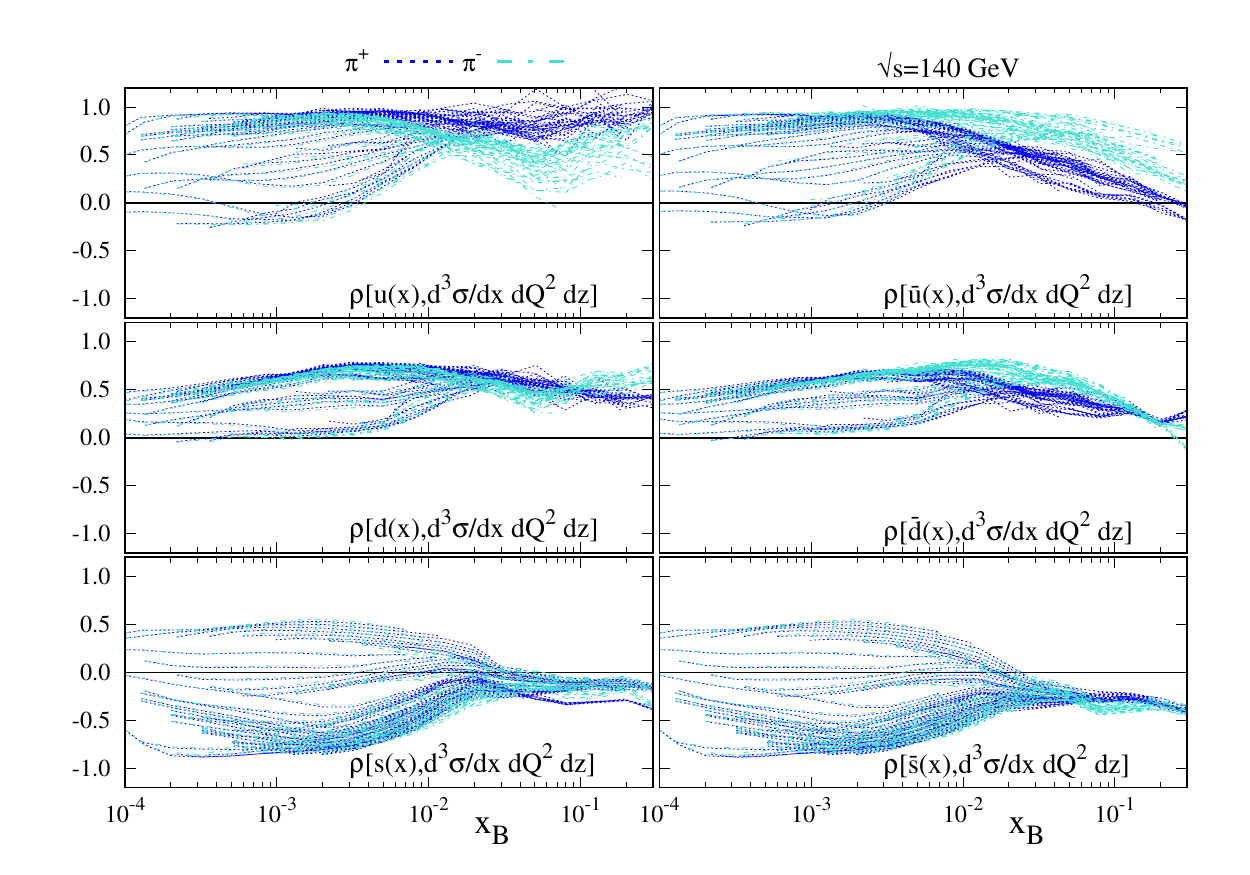,width=0.9\textwidth}
\epsfig{figure=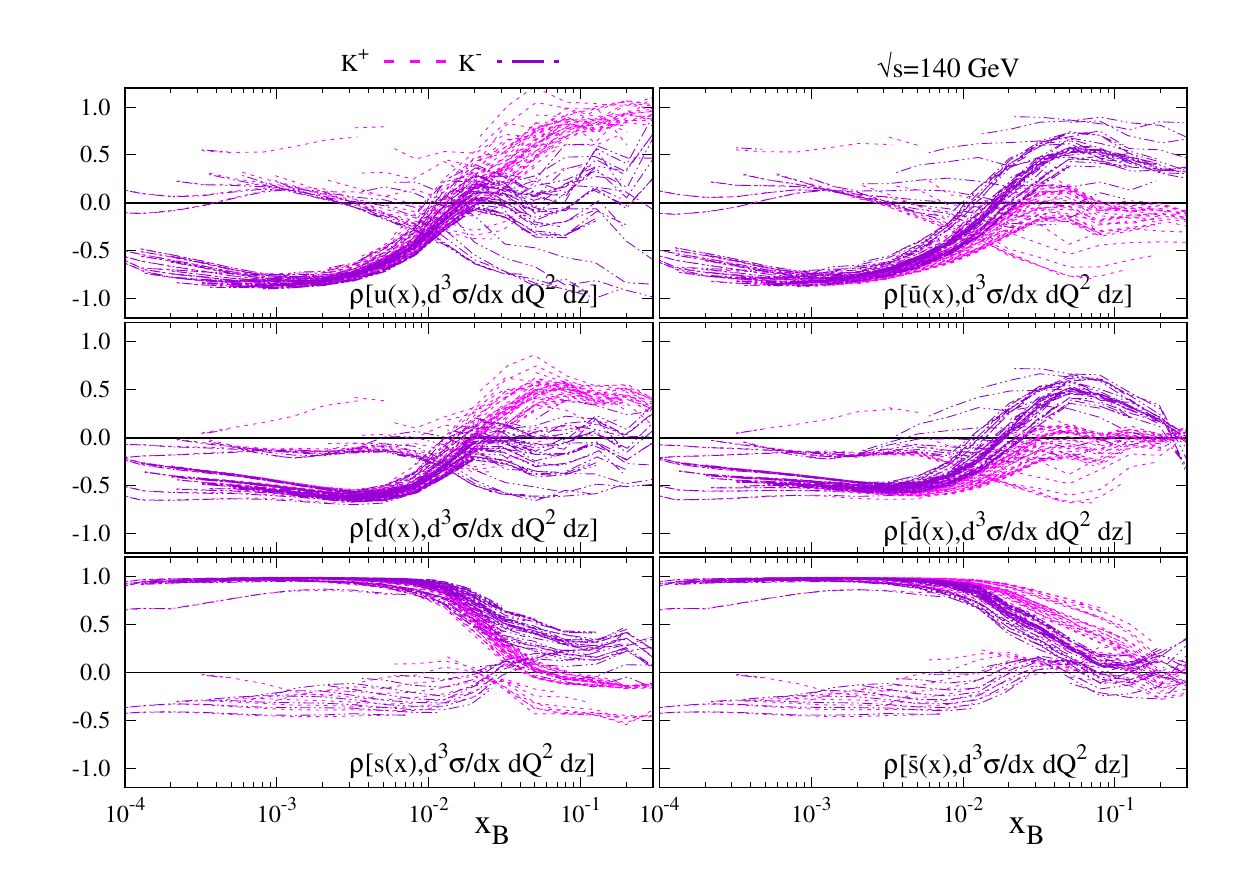,width=0.9\textwidth}
\caption{Correlation coefficient $\rho$ between the charged kaon (magenta and violet) and pion 
(cyan and blue) production in SIDIS at an EIC, and the light-quark PDFs, as a function of $x_{B}$ 
at $\sqrt{s} = 140$~GeV. Each box in the figures represents the correlation with one 
specific quark flavour. Each line corresponds to a different bin in $Q^{2}$ and $z$.}
\label{fig:rho_20x250}
\end{figure*} 
%
\begin{figure*}[h]
\epsfig{figure=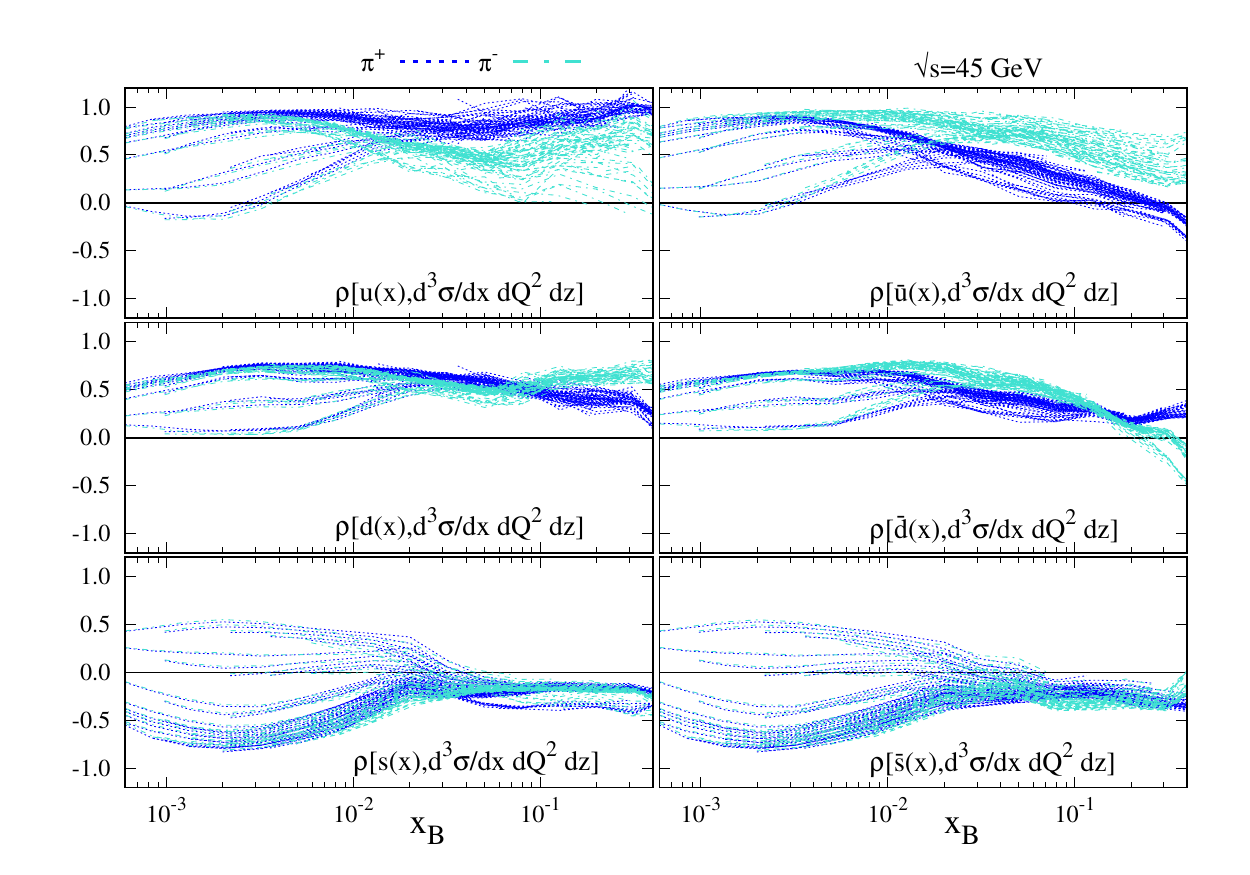,width=0.9\textwidth}
\epsfig{figure=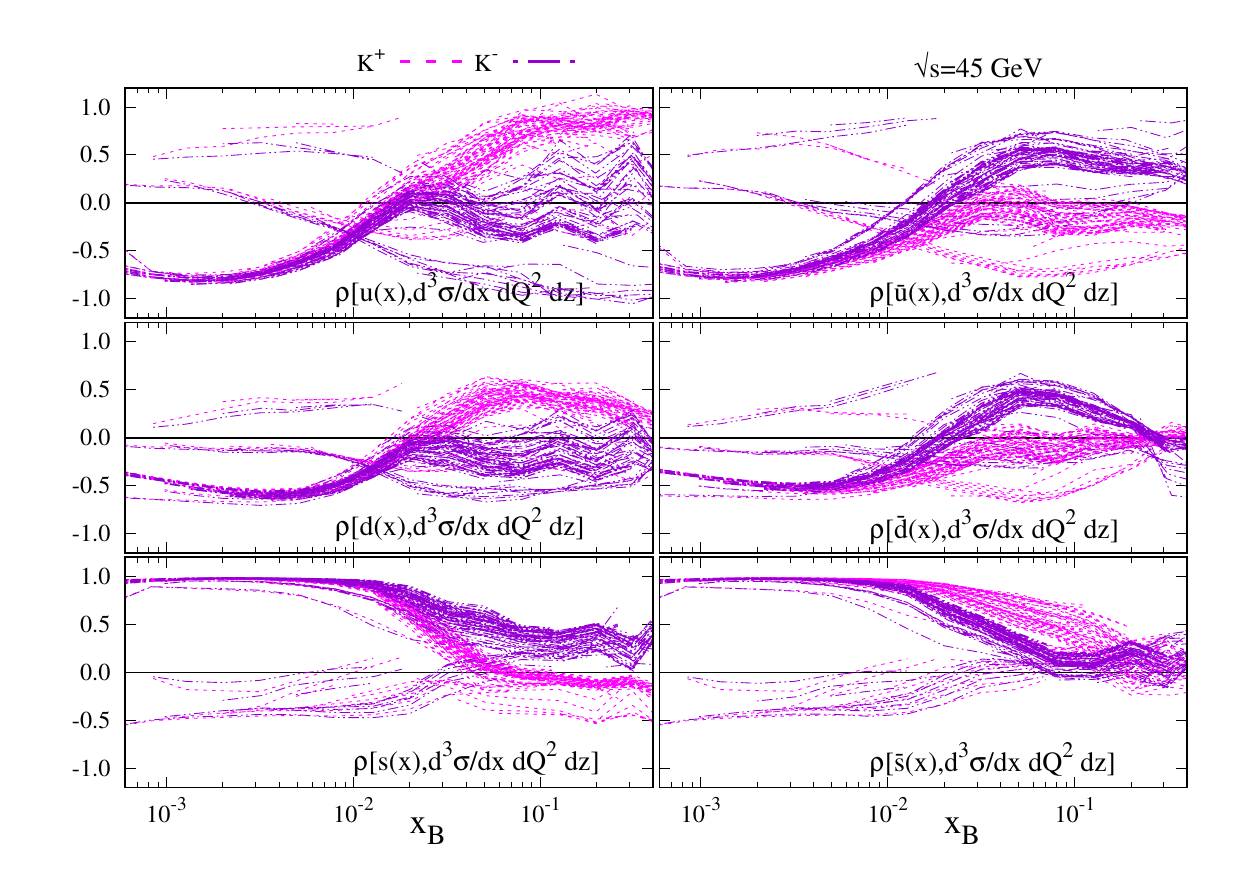,width=0.9\textwidth}
\caption{Same as Fig.~\ref{fig:rho_20x250}, but for $\sqrt{s} = 45$~GeV.}
\label{fig:rho_5x100}
\end{figure*} 
%
In this section, we present the results for the correlation and sensitivity coefficients 
between pion and kaon pseudodata and the non-perturbative distributions (PDFs and FFs), 
assessing the regions of phase space where the data have the largest impact on the 
determination of these distributions. We also assess the impact of two different c.m.s. 
energies ($\sqrt{s} = 45$~GeV and $140$~GeV) of the future EIC.

Figs.~\ref{fig:rho_20x250} and \ref{fig:rho_5x100} show the correlation coefficients between 
PDFs for different quark flavours and the SIDIS cross sections for charged pions and kaons 
as a function of $x_{B}$ for c.m.s. energies $\sqrt{s} = 140$~GeV and $\sqrt{s} = 45$~GeV, 
respectively. 
The coefficients for $\pi^{+}$ and $\pi^{-}$ are represented by the dotted (blue) and 
dashed-dotted (light-blue) lines respectively, while those for $K^{+}$ and $K^{-}$ are shown 
as the dashed (pink) and  the long dashed-dotted (violet) lines, respectively.  
The correlation coefficients are calculated for the kinematics $\{x_B,Q^{2},z\}$ of each 
pseudodata point, evolving the PDFs to the adequate $\{x_B,Q^{2}\}$, while the
lines interpolate between data points at the same $\{z,Q^{2}\}$.

As expected, larger correlations are typically found for quark flavours that are valence-like
for the final-state hadron, e.g., $\bar{d}$ in $\pi^{+}$, in the region of $x_B$ where such 
flavour is most abundant in the proton target. 
The valence flavours show larger correlations at larger $x_{B}$, e.g., at larger 
$x_{B}$, $\pi^{+}$ ($\pi^{-}$) production cross sections show a stronger correlation 
with $u\,(\bar{u})$ and $\bar{d}\,(d)$ quark distributions, while the ones with 
$s$ and $\bar{s}$ quarks are suppressed. 
For lower values of $x_{B}$ the data probe sea-quark distributions, 
for which $s\sim u\sim d$ and $q=\bar{q}$, balancing the correlation coefficients of 
$\pi^{+}$ and $\pi^{-}$ and enhancing the anti-correlations with strange quarks, 
which become of the same magnitude as the light quarks.  

In the case of $u$ quarks, the correlation coefficient for the simulated cross 
section for positively charged pions is close to one for the full range of $x_{B}$ probed, 
while the same holds for $\bar{u}$ and the cross section for negatively charged pions, as is 
foreseeable considering that $D_{u}^{\pi^{+}}=D_{\bar{u}}^{\pi^{-}}$. Ultimately, most of 
the constraints for these distributions will therefore come from the pion production data. 
It is also worth noticing that due to the electric-charge factors, the correlation 
coefficients for the (anti-)up quark distribution are enhanced compared to those of 
the (anti-)down quark distribution.

Similar features can be found for the kaon production cross section. In this case, stronger 
correlations are obtained for the $u$ ($\bar{u}$) and $\bar{s}$ ($s$) quarks, in agreement 
with the $K^{+}$($K^{-}$) valence composition.  
For values $x_{B}>10^{-2}$, the correlation with $\bar{s}\,(s)$ almost vanishes, as the 
data probe mainly the proton's valence distributions for these values of $x_{B}$, 
while the proton only has sea strange quarks. 
For lower values of $x_{B}$, one can access the (anti)strange-quark distributions.
In this $x_{B}$ range, the correlation coefficients for these distributions get close to one, 
while some anti-correlation is obtained with (anti-)up and (anti-)down quarks. 
Comparing the correlation coefficients, it can be anticipated that the 
constraint on the strange content of the proton will essentially come from the kaon data 
at small $x_{B}$. 
The results seem to indicate that kaon data could also be relevant for 
the determination of the (anti-)up quark distributions at higher values of $x_{B}$. 
However, as will be discussed later in this section, for higher $x_B$, these data become 
less relevant.

Regarding the different energy configurations, it is worthwhile noticing that while the 
correlation coefficients obtained for the lower energy configuration span slightly higher 
values of $x_{B}$ than those obtained for the higher energy configuration, 
the correlations do not show significantly different features. 

\begin{figure*}[ht!]
\epsfig{figure=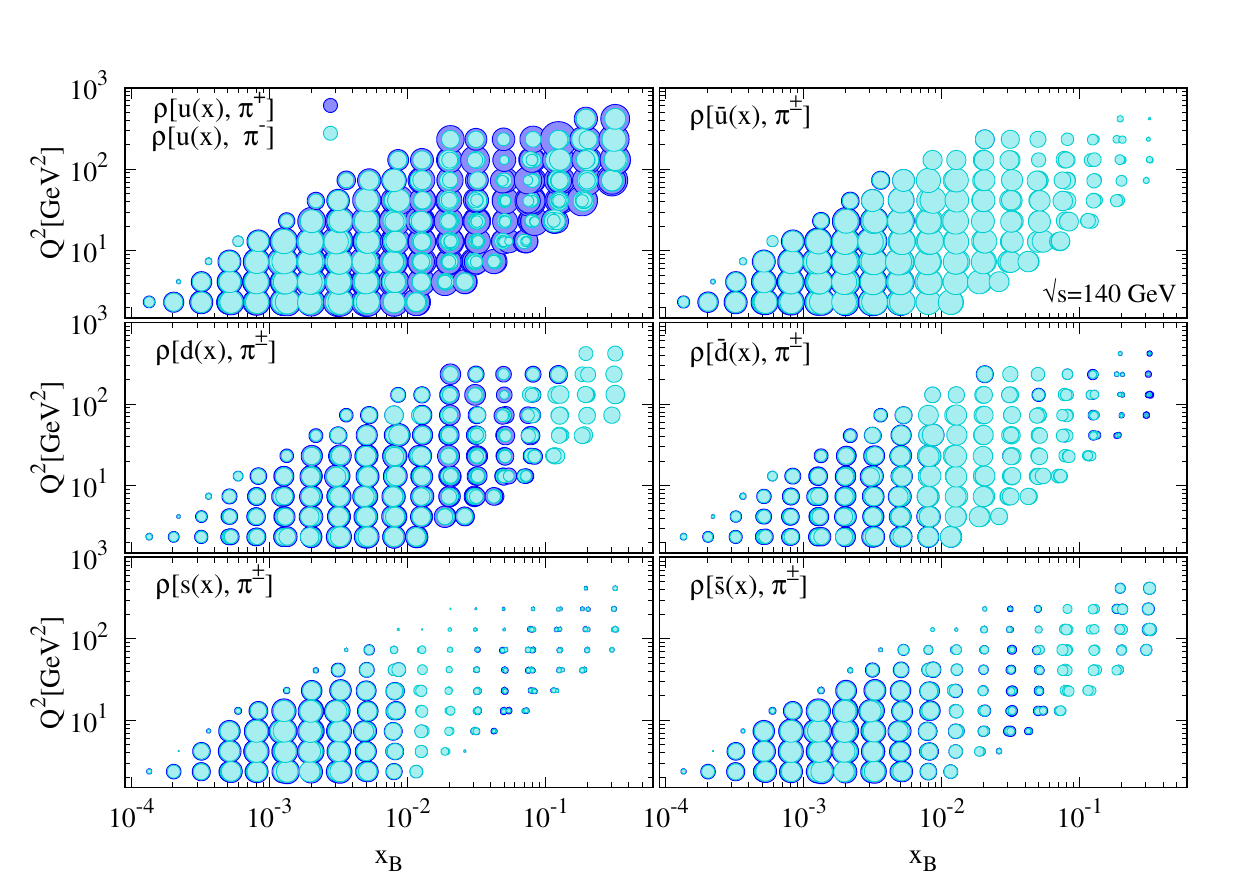,width=0.9\textwidth}
\caption{Correlation coefficient $\rho$ between the cross section for charged-pion production 
and the PDFs of the light quarks, as a function of both the Bjorken variable $x_{B}$, and the 
square of the momentum transfer $Q^{2}$. For each data point, a circle of which the radius 
represents the correlation is depicted.}
\label{fig:corr_map_pion_20x250}
%
\epsfig{figure=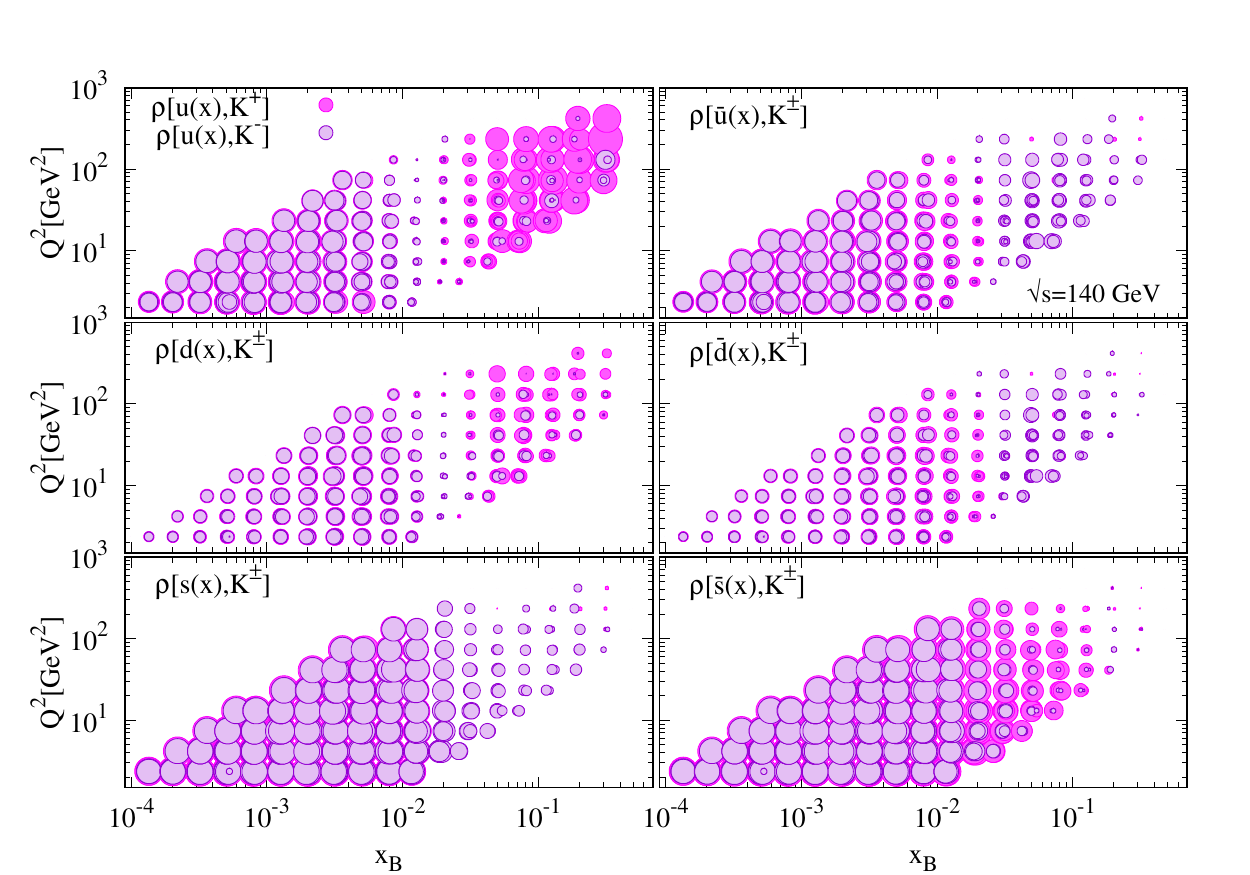,width=0.9\textwidth}
\caption{Same as Fig.~\ref{fig:corr_map_pion_20x250} for charged-kaon production.}
\label{fig:corr_map_kaon_20x250}
\end{figure*} 
%
\begin{figure*}[t!]
\epsfig{figure=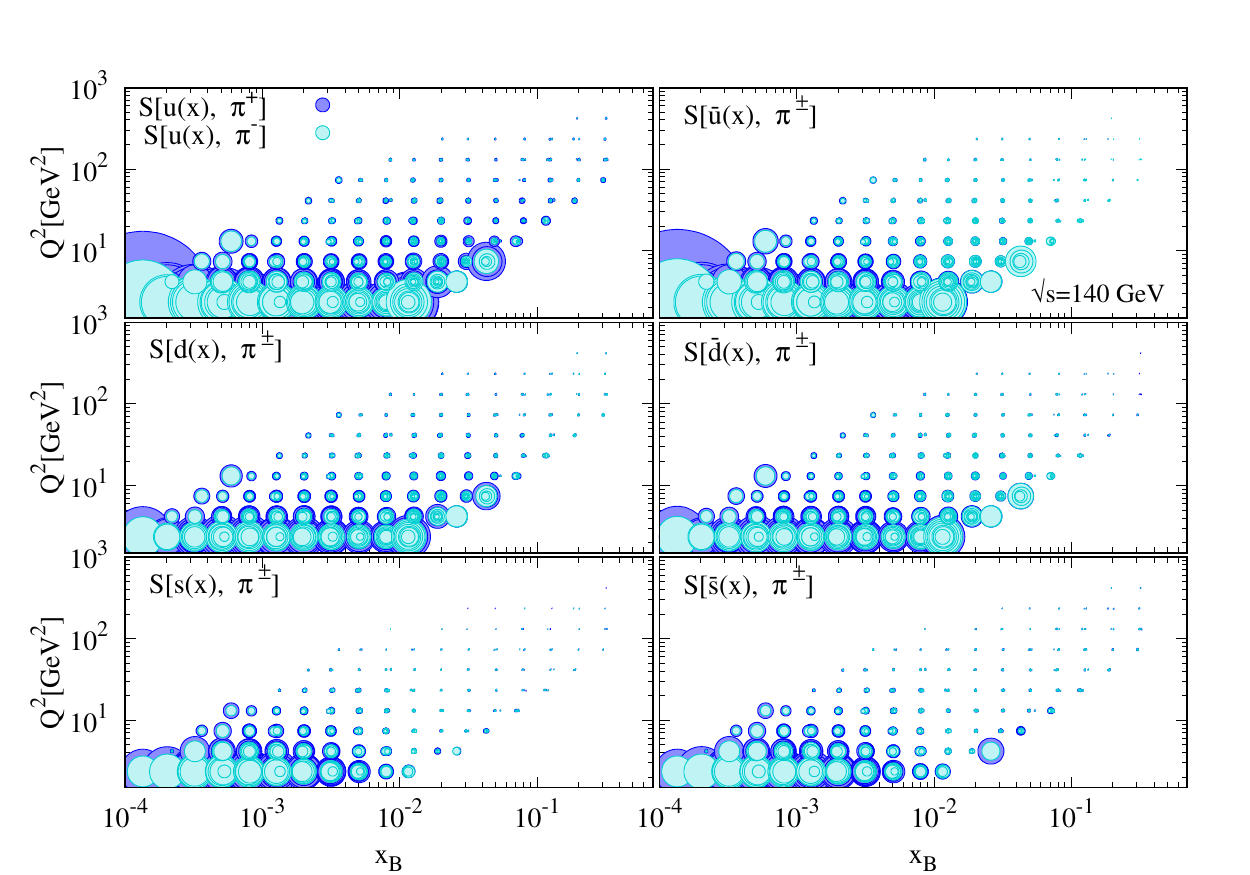,width=0.9\textwidth}
\vspace*{-0.5cm}
\caption{Sensitivity coefficients $S$ between the cross section for charged-pion production, 
and the different light-quark parton distributions, as a function of $x_{B}$ and the 
transferred momentum squared $Q^{2}$. As in \ref{fig:corr_map_pion_20x250}, each circle 
corresponds to a particular kinematic configuration $\{x_{B},Q^{2},z\}$ associated with a 
point from the pseudodata. Its radius corresponds to the value of the sensitivity coefficient 
for that particular configuration.} 
\label{fig:corr_map_pion_20x250_esc}
%
\epsfig{figure=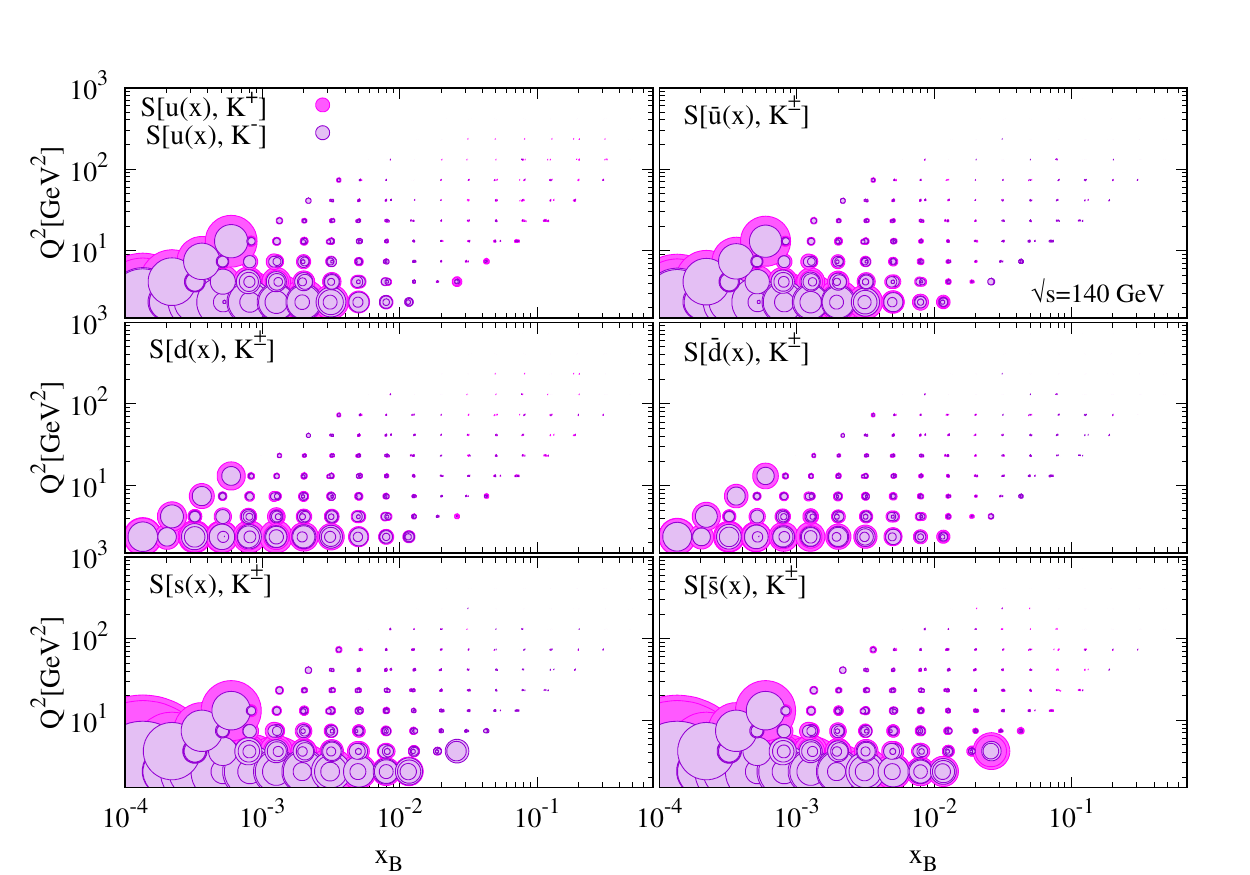,width=0.9\textwidth}
\caption{Same as Fig.~\ref{fig:corr_map_pion_20x250_esc} for charged-kaon production.}
\label{fig:corr_map_kaon_20x250_esc}
\end{figure*} 
%

In order to have a better insight into which data sets constrain best the PDFs, it is illustrative
to plot the correlation coefficients as a function of both $x_{B}$ and $Q^{2}$. In 
Figs.~\ref{fig:corr_map_pion_20x250} and \ref{fig:corr_map_kaon_20x250}, we show the 
correlation coefficients for pion and kaon production in SIDIS, and the parton
distribution for the different light quarks over the $x-Q^2$ plane. For each pseudodata point, 
we plot a circle with a radius proportional to the absolute value of correlation coefficient. 
Similar considerations on the correlations to those discussed for Figs. \ref{fig:rho_20x250} and 
\ref{fig:rho_5x100} hold here, however, now the $Q^2$ dependence of the correlations is made 
explicit. Notice that larger values of $Q^2$ are correlated to larger values of $x_{B}$, as usual
for DIS experiments. For these values, a clear hierarchy emerges with the largest correlation coefficients
for quark flavours that are {\em valence}-like for the final-state hadron, have the largest
charge ($e_{q}=2/3$) factor, and are {\em valence}-like also in the proton target. 
The weakest correlation is, as expected, for the strange quarks and pions at larger 
$x_{B}$ and $Q^2$. However, the full strength of this kind of plot will become more apparent when 
studying the sensitivity coefficients.

As discussed in Section \ref{sec:correlation}, the correlation coefficients only give an estimate 
of the \textit{potential} impact that a new data set could have if included in a new global fit, 
because the experimental precision of the data is not taken into account, and more specifically,
because the correlation coefficients do not describe how precise the new data are compared to those used for the PDF determination 
nor how well the new data is described by the existing PDFs, within their uncertainty.
In this respect it is more instructive to examine the sensitivity, or weighted correlations, 
defined in Section \ref{sec:correlation}. 
In Figs.~\ref{fig:corr_map_pion_20x250_esc} and \ref{fig:corr_map_kaon_20x250_esc}, we show the 
sensitivity coefficient as a function of both $x_{B}$ and $Q^{2}$. 
As before, the size of the circle for each data point is determined by the absolute value of the 
sensitivity coefficient. Notice that contrary to the correlation coefficients, the 
sensitivity coefficients are not normalised to unity, but instead they are proportional to the ratio between the uncertainty 
in the cross section propagated from the PDFs and that coming from the measurement.

Comparing Fig.~\ref{fig:corr_map_pion_20x250} with Fig.~\ref{fig:corr_map_pion_20x250_esc}, 
and  Fig.~\ref{fig:corr_map_kaon_20x250} with Fig.~\ref{fig:corr_map_kaon_20x250_esc}, it 
becomes evident that the most significant impact on the PDFs is expected to come from the 
low $x_{B}$ region, which for SIDIS like for DIS is associated to the low $Q^{2}$ region. 
Even though the charged-hadron--production data have high correlations with different parton 
distributions throughout the complete kinematic range covered, the most important impact 
is expected for $x_{B}<10^{-2}$ and $Q^{2}<10^{2}$, since for higher $x_{B}$ and $Q^2$, 
the PDFs are already well constrained.

\begin{figure}[ht!]
\epsfig{figure=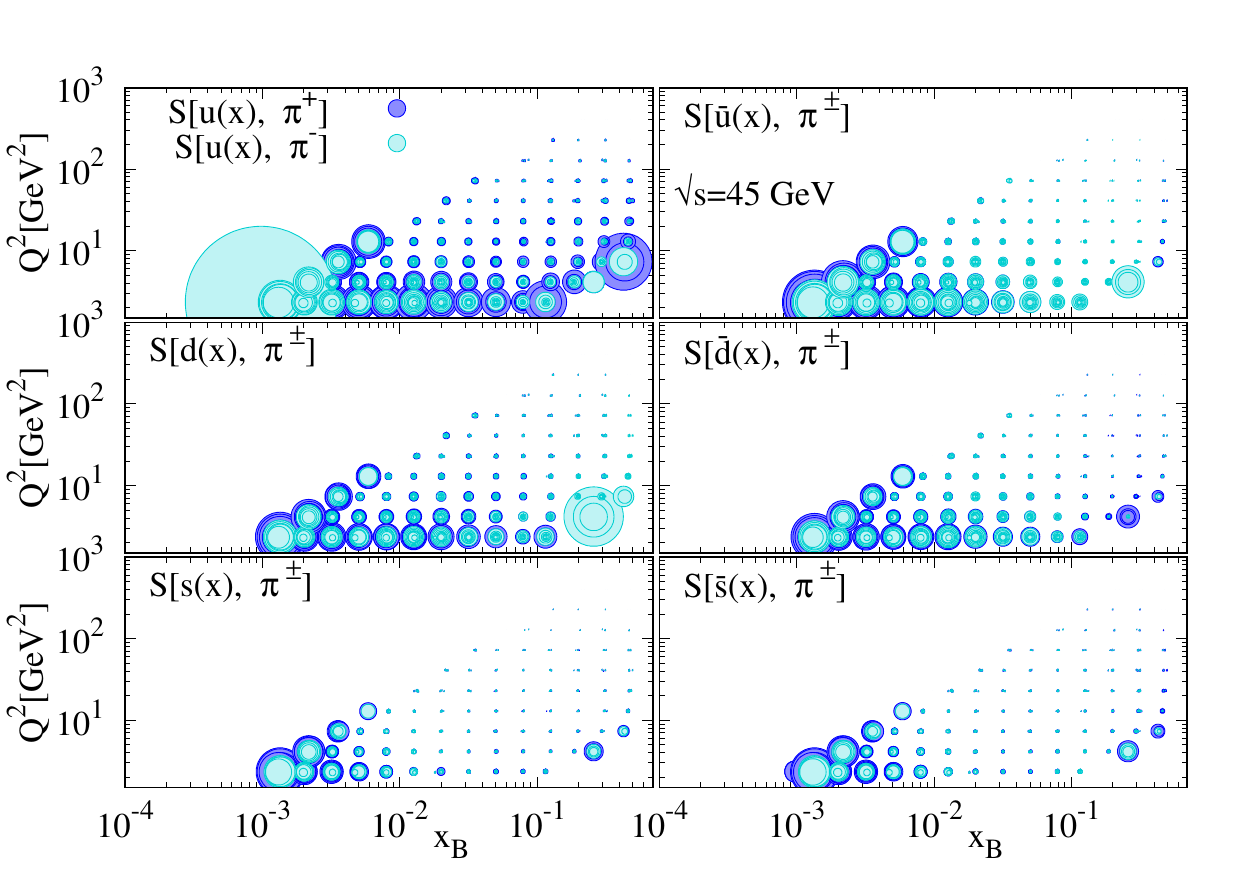,width=0.525\textwidth}
\caption{Same as Fig.~\ref{fig:corr_map_pion_20x250_esc}, for a c.m.s. energy $\sqrt{s} = 45$~GeV. 
To make the comparison clear, we keep the same scales as in the previous plots.} 
\label{fig:corr_map_pion_5x100_esc}
\end{figure}

\begin{figure}[t!]
\epsfig{figure=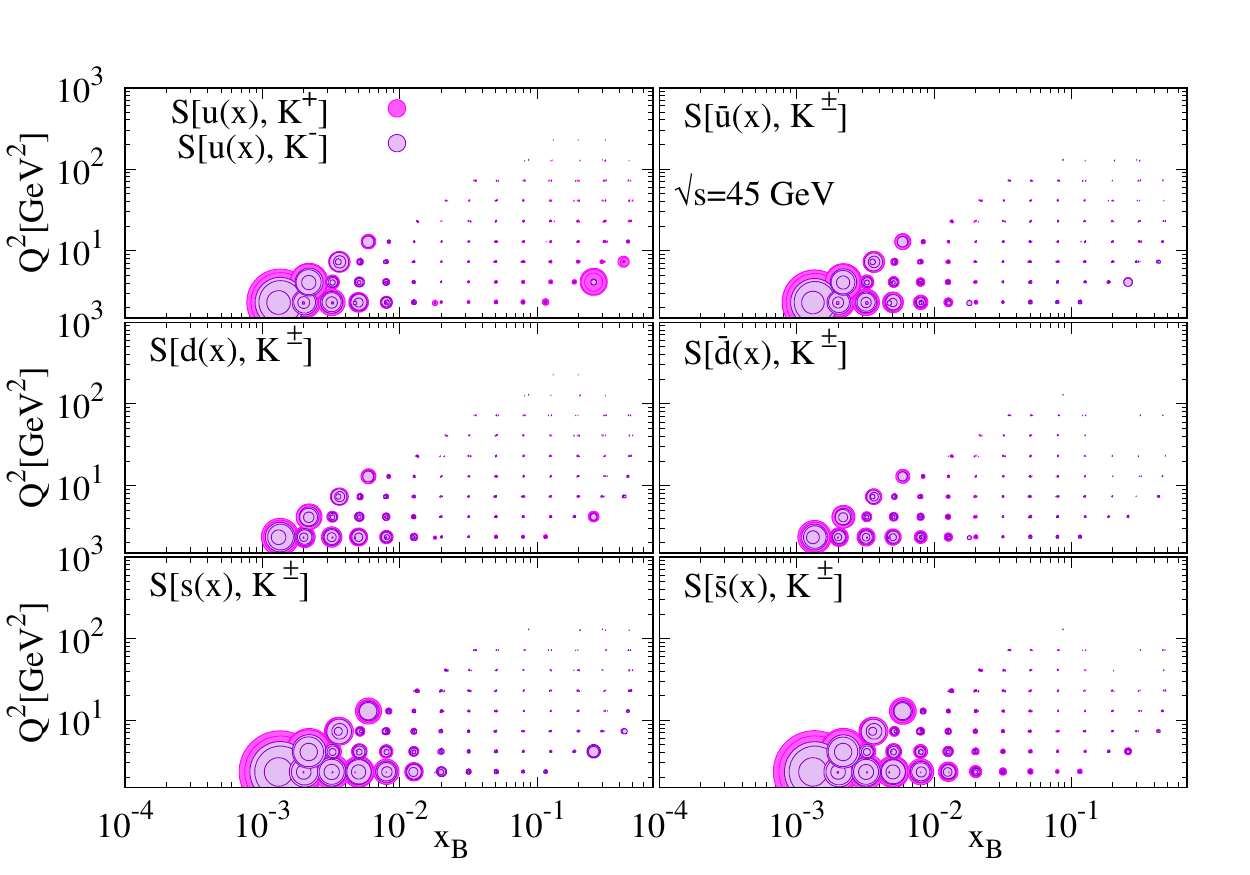,width=0.525\textwidth}
\caption{Same as Fig.~\ref{fig:corr_map_pion_5x100_esc} for charged-kaon production.}
\label{fig:corr_map_kaon_5x100_esc}
\end{figure}

At this point, it is also enlightening to compare the sensitivity estimates obtained for 
$\sqrt{s} = 140\,\text{GeV}$ with those for $\sqrt{s} = 45\,\text{GeV}$. 
The latter are shown in Figs.~\ref{fig:corr_map_pion_5x100_esc} 
and \ref{fig:corr_map_kaon_5x100_esc} for pions and kaons, respectively. 
For this lower c.m.s. energy, the impact of the SIDIS data is restricted to the kinematic 
region given by $10^{-3}<x_{B}<10^{-2}$, 
where the highest values of sensitivity are obtained. However, in spite of the high correlations 
of the cross sections at higher values of momentum fractions, the expected impact is diluted 
by the relative error. Notice that for this energy configuration, the most sensitive region explored with the  $\sqrt{s} = 140\,\text{GeV}$ c.m.s configuration, i.e.,
the one shown to have the greatest constraining power in Figs.~\ref{fig:corr_map_pion_20x250_esc} 
and \ref{fig:corr_map_kaon_20x250_esc}, $10^{-4}<x_{B}<10^{-3}$ is not probed.

%

Regarding the impact that EIC SIDIS data could have in the extraction of fragmentation 
functions, it is worth noting that SIDIS data have a central role in global fits, since 
they provide almost all the separation between quark and antiquark fragmentation and a 
good deal of that between flavours. The remarkably precise data from inclusive 
single-hadron production in 
electron-positron annihilation (SIA) is mostly sensitive to the singlet combination of 
fragmentation functions, while hadron production in proton-proton collisions mainly probes 
gluon fragmentation. As explained in Section \ref{sec:correlation},  the correlation 
and sensitivity coefficients can also be defined within the improved Hessian approach, 
considering the variations of the observables over the hessian eigenvector sets, which is 
the technique implemented in the charge and flavour discriminated DSS extractions of FFs 
and their updates \cite{deFlorian:2007aj,deFlorian:2014xna,deFlorian:2017lwf}. 

\begin{figure*}[ht!]
\epsfig{figure=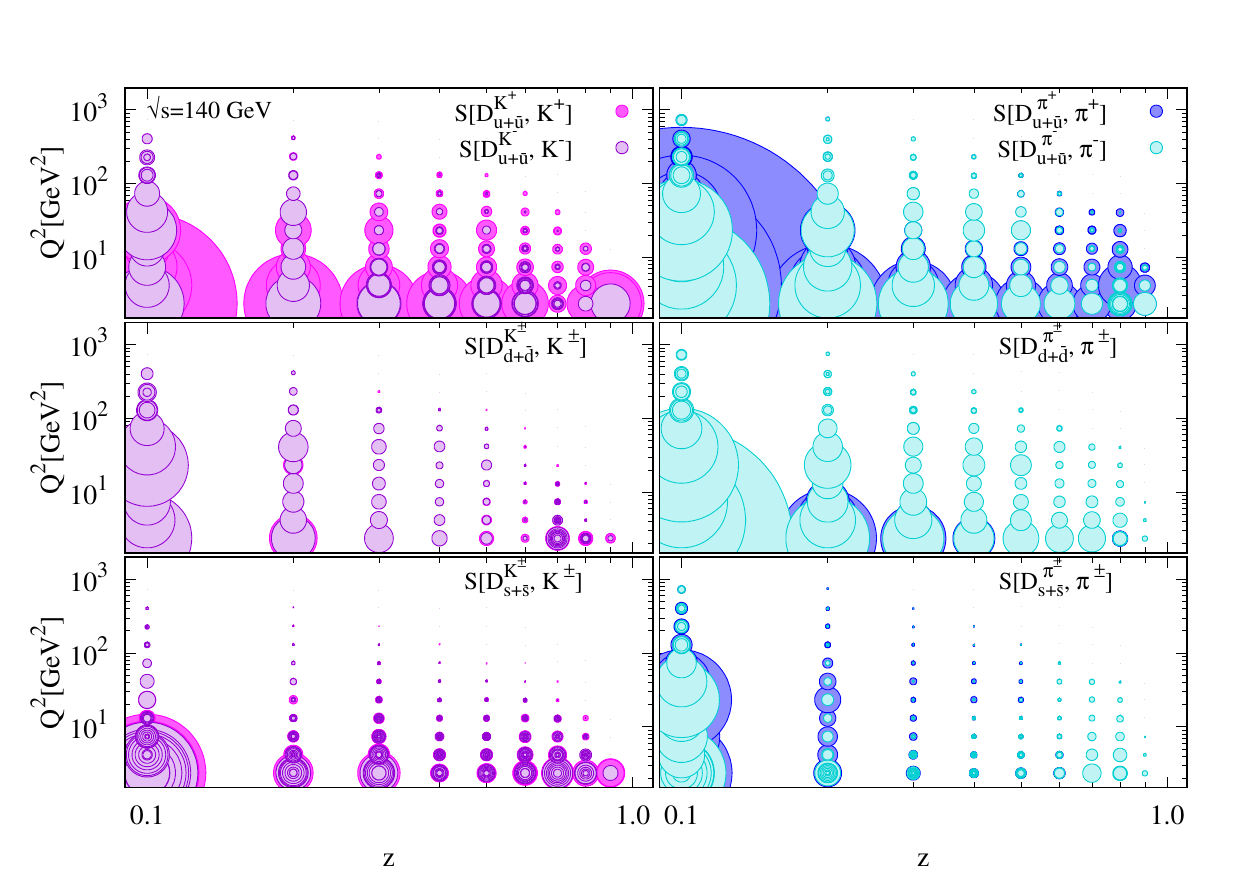,width=0.9\textwidth}
\caption{Sensitivity coefficients between the cross section for charged-hadron production at 
$\sqrt{s}=140$ GeV, pions (blue and light-blue) and kaons (pink and violet), and the singlet FF combination 
$D^{H^{\pm}}_{q+\bar{q}}$ for the different light quark flavours. The coefficients are 
obtained using the Hessian formalism described in Sec.~\ref{sec:correlation} (see text).}
\label{fig:corr_map_ffsum_20x250_esc}
%
\vspace*{-0.2cm}
\epsfig{figure=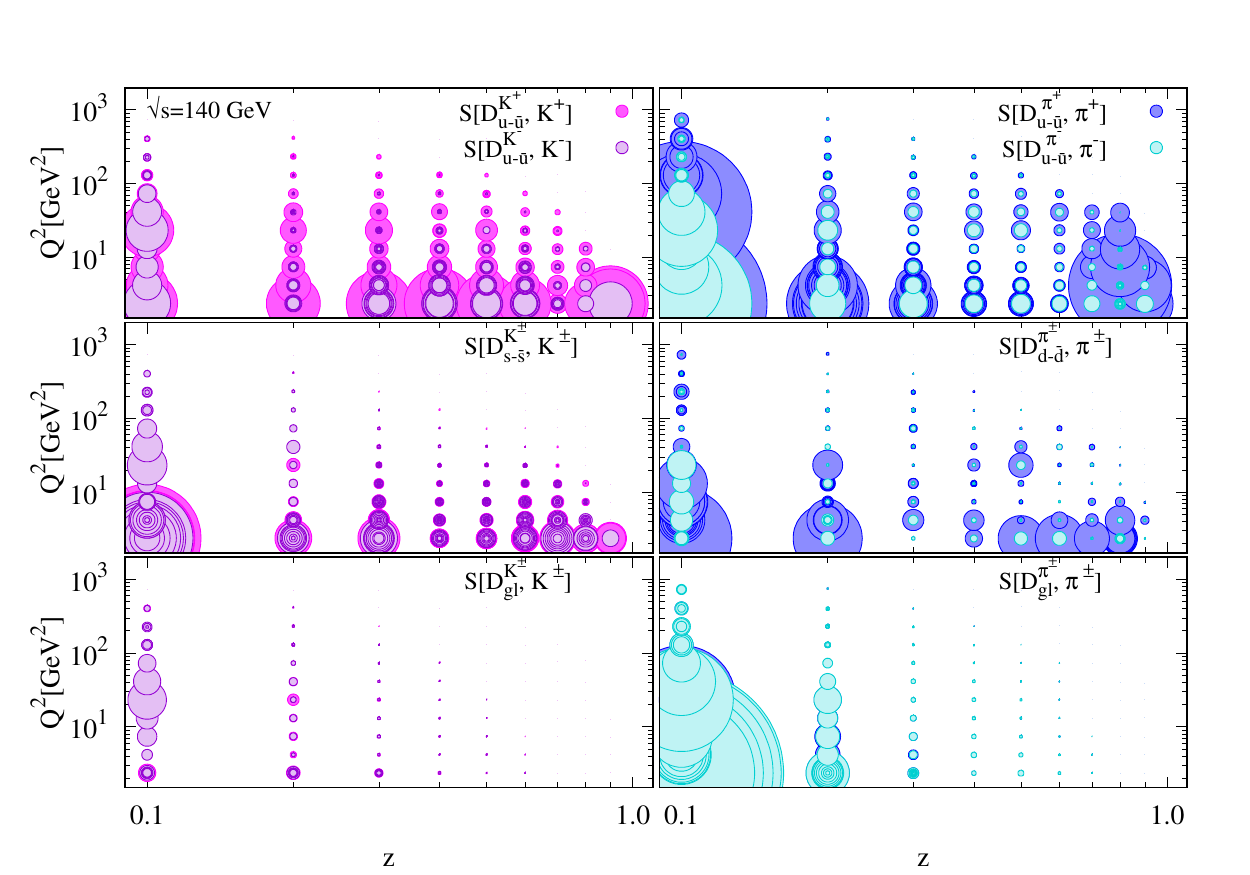,width=0.9\textwidth}
\caption{Same as Fig.~\ref{fig:corr_map_ffsum_20x250_esc} for the non-singlet combination 
of FFs $D^{H^{\pm}}_{q-\bar{q}}$.}
\label{fig:corr_map_ffrest_20x250_esc}
\end{figure*}

In order to establish the kinematic regions where the EIC SIDIS data could have the most 
significant impact for FFs, we compute the sensitivity coefficients between the cross section 
for charged pion and kaon production and the {\em plus} and {\em minus} combinations 
$D^{H^{\pm}}_{q+\bar{q}}$ and $D^{H^{\pm}}_{q-\bar{q}}$ discriminating for each final-state
hadron, and for each of the light-quark flavours. The former are the combinations expected 
to be constrained by SIA while the latter are better constrained by SIDIS. 
In Figs.~\ref{fig:corr_map_ffsum_20x250_esc} and~\ref{fig:corr_map_ffrest_20x250_esc}
we show the sensitivities as a function of $z$ and $Q^{2}$. The left panels correspond to 
the coefficients for the cross sections for kaon production, while the right panels are 
associated with the cross sections for pion production. The coefficients are calculated for 
the c.m.s. energy of $\sqrt{s} =  140$ GeV. 

%
%
%
As can be seen in Figs.~\ref{fig:corr_map_ffsum_20x250_esc} and~\ref{fig:corr_map_ffrest_20x250_esc}, 
the sensitivities typically grow as $Q^{2}$ and $z$ decrease, mainly because of the 
FF uncertainties, which increase in these limits, and are non-negligible for both the 
{\em plus} and {\em minus} combinations, suggesting a significant constraining power not 
only for the charge separation, but also competitive for discriminating between quark 
flavours. Notice that in the DSS FF extractions \cite{deFlorian:2014xna,deFlorian:2017lwf}, 
the only data below $z\sim 0.1$ come from LEP experiments, at very high energy scales, which
explains the impressive increase of the sensitivity.

For completeness, we also include in Fig.~\ref{fig:corr_map_ffrest_20x250_esc} 
the results for the sensitivity coefficients between the gluon FF $D^{H^{\pm}}_{g}$ and 
the charged hadron production cross sections. These are found to be marginal, since the 
constraining power of SIDIS data is not competitive with the RHIC and ALICE proton-proton 
collision data already included in the global fits, except below $z \simeq 0.2$.
We do not show the correlations for $D^{K^{\pm}}_{d-\bar{d}}$ and $D^{\pi^{\pm}}_{s-\bar{s}}$, 
since these combinations are assumed to vanish in the DSS sets because of flavour symmetry 
considerations.

\subsection{Results for the reweighting using EIC SIDIS pseudodata.}\label{subsec:reweighting}
%
While the correlation and sensitivity coefficients are very useful tools to anticipate and
identify the kinematic regions where a given data set can be most relevant for constraining 
parton densities or fragmentation functions, ultimately the effect of the inclusion of the 
new data on the distributions need to be explicitly assessed by performing new global 
fits or a reweighting of a set of replicas. In this section we present and discuss the 
results of the reweighting exercise performed using the pseudodata generated for charged 
pion and kaon production in SIDIS described in Section \ref{sec:status}, and show 
the resulting set of modified PDFs and FFs as well as combinations of these distributions 
that quantify the degree of the charge and flavour symmetry breaking.

\begin{figure*}[ht!]
\vspace*{0.2cm}
\hspace*{-0.1cm}
\epsfig{figure=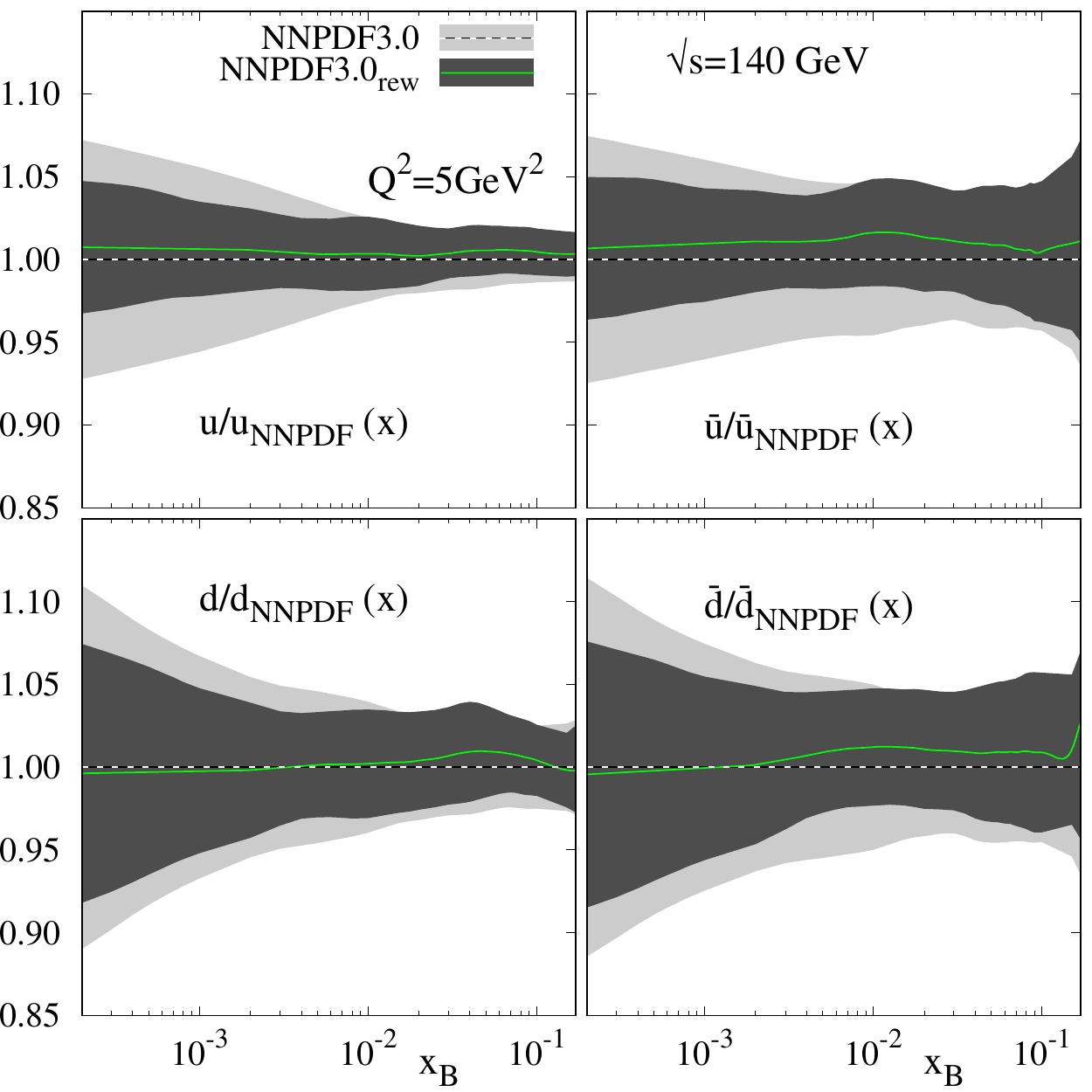,width=0.49\textwidth}
\epsfig{figure=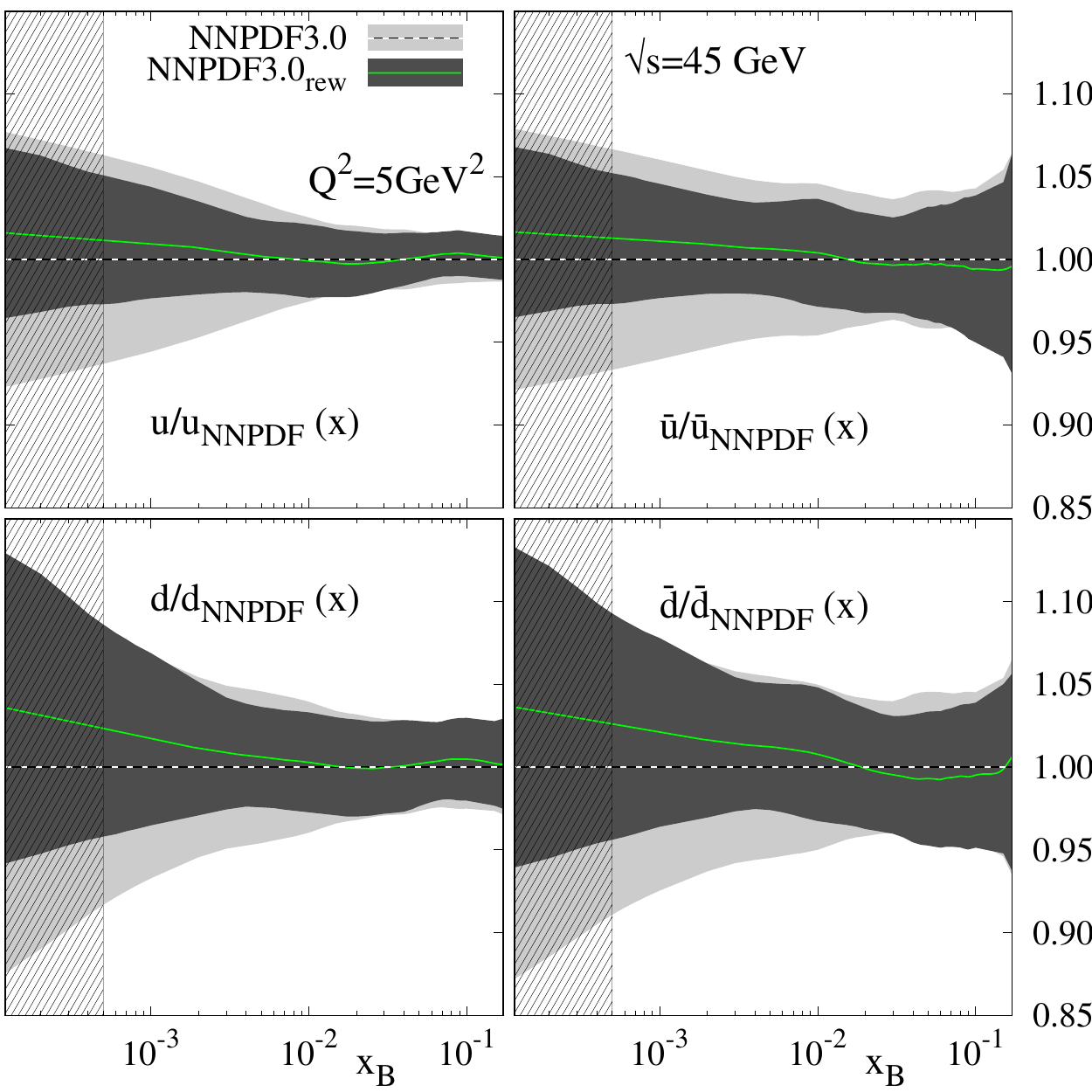,width=0.49\textwidth}
\caption{Reweighting of NNPDF3.0 NLO replicas for the $u$ and $\bar{u}$ quark distribution 
(upper panels) and $d$ and $\bar{d}$ quark distribution (lower panels) with EIC pseudodata. 
The four panels on the left-hand side correspond to $\sqrt{s} = 140$ GeV pseudodata, while those 
on the right-hand side are for $\sqrt{s} = 45$ GeV. The shaded area is the region of $x_{B}$ 
not covered by the latter energy configuration.
The distributions are normalised to the NNPDF3.0 best fit. The solid (green) lines and dark grey 
bands represent the results for the distributions after the reweighting procedure and the 
corresponding uncertainty bands, respectively. All results are shown at a scale of 
$Q^{2}=5$~GeV$^2$.\label{fig:u+d}}
%
\vspace*{0.91cm}
\hspace*{-0.1cm}
\epsfig{figure=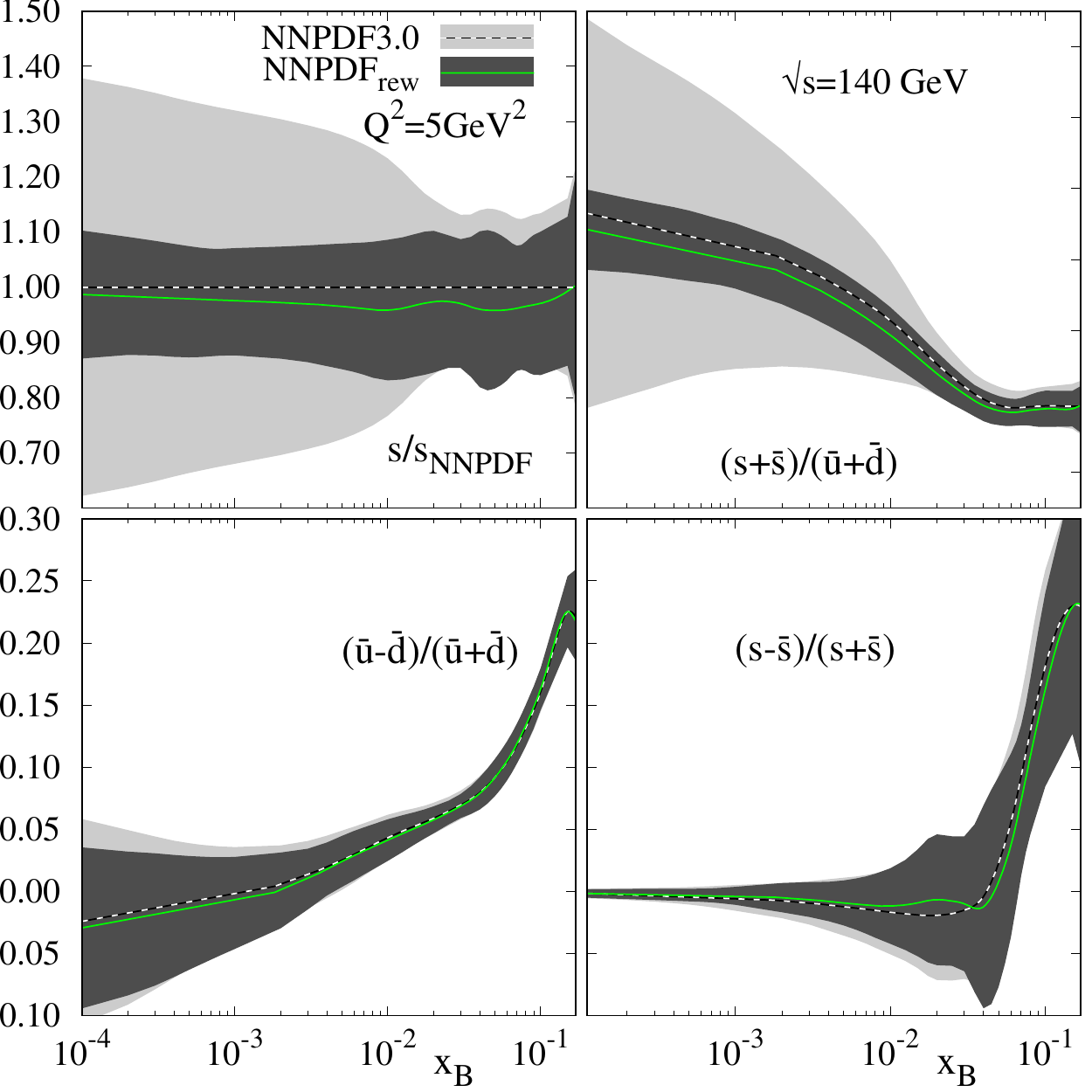,width=0.49\textwidth}
\epsfig{figure=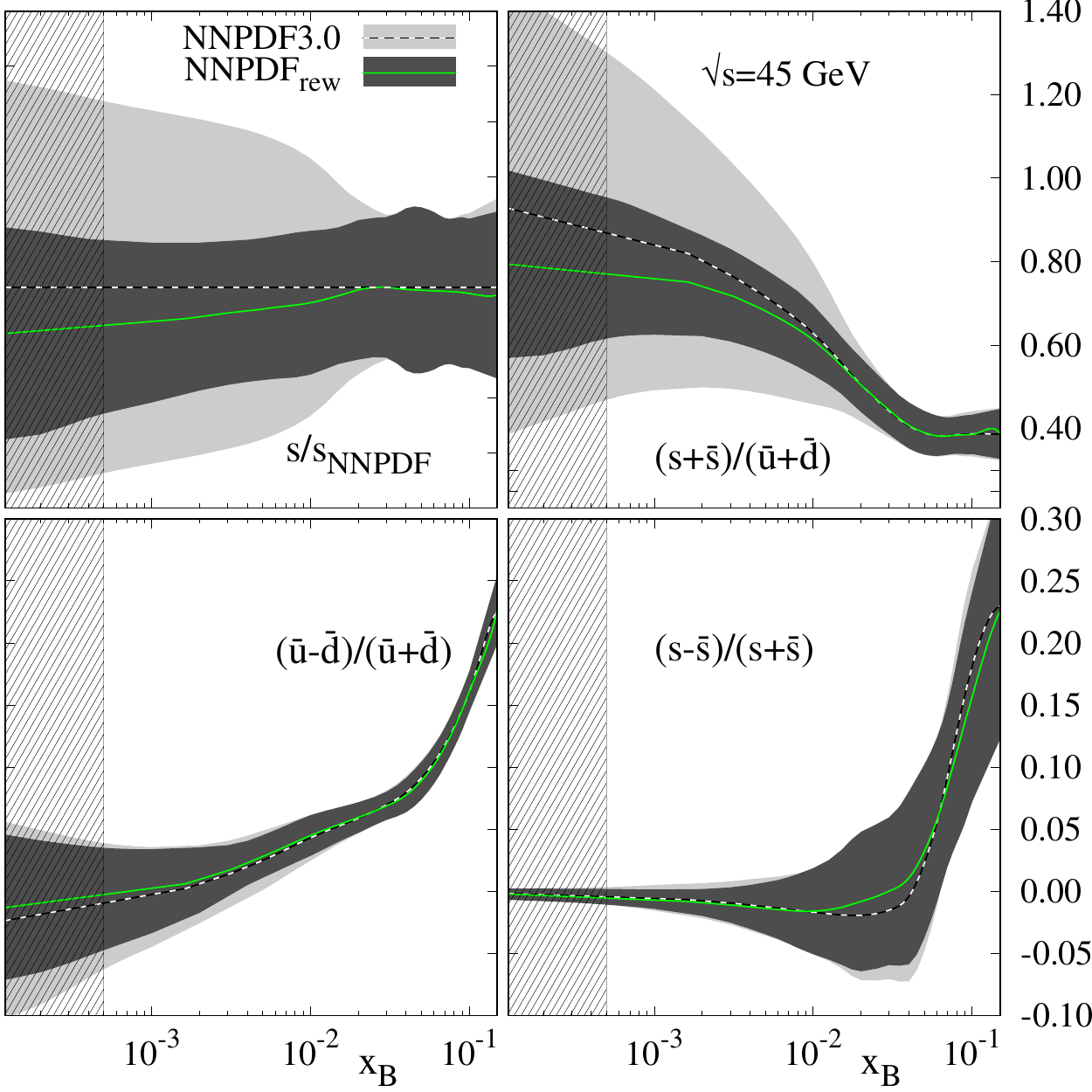,width=0.49\textwidth}
\caption{The same as Fig.\ref{fig:u+d}, but for the strange quark distribution (upper
panels) and for the PDF combinations sensitive to charge and isospin (lower panels) 
symmetry breaking. Again, the results are shown at a scale of $Q^{2}=5$~GeV$^2$ and 
are normalised to the NNPDF3.0 best fit.
\label{fig:s+asymmetry}}
\vspace*{0.5cm}
\end{figure*}

We start with the non-strange light-quark PDFs.  In Fig.~\ref{fig:u+d} we show the effect
of reweighting a set of 1000 PDF replicas of the NLO NNPDF3.0 set with EIC SIDIS pseudodata. 
The four panels on the left-hand side correspond to a set of pseudodata at
a c.m.s. energy of $\sqrt{s} = 140$ GeV, while those on the right-hand side, to 
$\sqrt{s} = 45$ GeV. In both cases, the number of effective replicas $N_{eff}$ is 
above 80, ensuring that the modified distributions are an accurate representation of 
the original probability distribution. 

In reweighting, pseudodata with $z<0.1$ is excluded, since the FFs used to compute 
the central values have rather large uncertainties that hinder any constraining effect. On 
the other hand, pseudodata points with $Q^{2}<2$ GeV$^2$ are also excluded from
the reweighting since their statistical power is so restrictive that the resulting 
number of effective 
replicas after the reweighting is extremely low ($N_{rep}\approx10$). Similarly, 
it should also be noted that the pseudodata coming from the two
alternative c.m.s configurations are not combined into a single reweighting, 
given that the constraints imposed by the whole data set leave a low number of effective replicas.
This indicates that if the whole data set were to be included in a global fit, the impact on the uncertainties would 
be stronger, but it would require either a new global fit or a reweighting with a much larger 
number of replicas. 

Since the pseudodata is generated around the NNPDF3.0 best fit result, the main effect on 
the distributions is expected to be a reduction on the uncertainty bands, with a very minor 
variation of the central values. Indeed, the new SIDIS information can at most balance small 
tensions already present between the data sets of the original fit.  The distributions and 
the corresponding uncertainty bands are normalised to the NNPDF3.0 best fit result, represented 
in the plots by the dashed (black) lines with light grey bands. The reweighted results are 
plotted as solid (green) lines with the dark grey uncertainty bands. The upper panels correspond 
to the $u$ and $\bar{u}$ quark distributions, 
while the lower panels show the analogous result for the $d$ and $\bar{d}$ quark. 

The most noticeable feature in the plots is the significant reduction in the uncertainty bands. 
The inclusion of the EIC pseudodata leads to a reduction of the uncertainty of order $30\%$ 
for the up quark, driven by the new kaon and pion data,  and $20\%$ for the down quark, 
led by the pion data. It is also worth noticing that the kinematic region 
where the impact of the SIDIS pseudodata is most important is precisely the region $x_{B}<10^{-2}$, 
as anticipated from the sensitivity coefficients calculation depicted in 
Fig.~\ref{fig:corr_map_pion_20x250_esc}. As stated in the previous section, 
in spite of the high correlation between the pion cross section and the (anti-)up 
quark distribution for higher values of $x_{B}$, the inclusion of the pseudodata through the 
reweighting procedure hardly modifies the distributions in that kinematic region. 
Indeed, while a smaller impact for the high $x_{B}$ region was expected according to the sensitivity 
coefficients, the fact that the distributions are hardly modified in that kinematic configuration 
is the result of the increasing uncertainty associated to the FFs. 
As mentioned in section \ref{sec:status}, the theoretical uncertainty coming from the FFs must 
be included in the reweighting procedure, thus attenuating the impact of the pseudodata in 
the regions where these uncertainties become larger than those of the PDFs. 

\begin{figure*}[ht!]
\vspace*{0.1cm}
\hspace*{-0.1cm}
\epsfig{figure=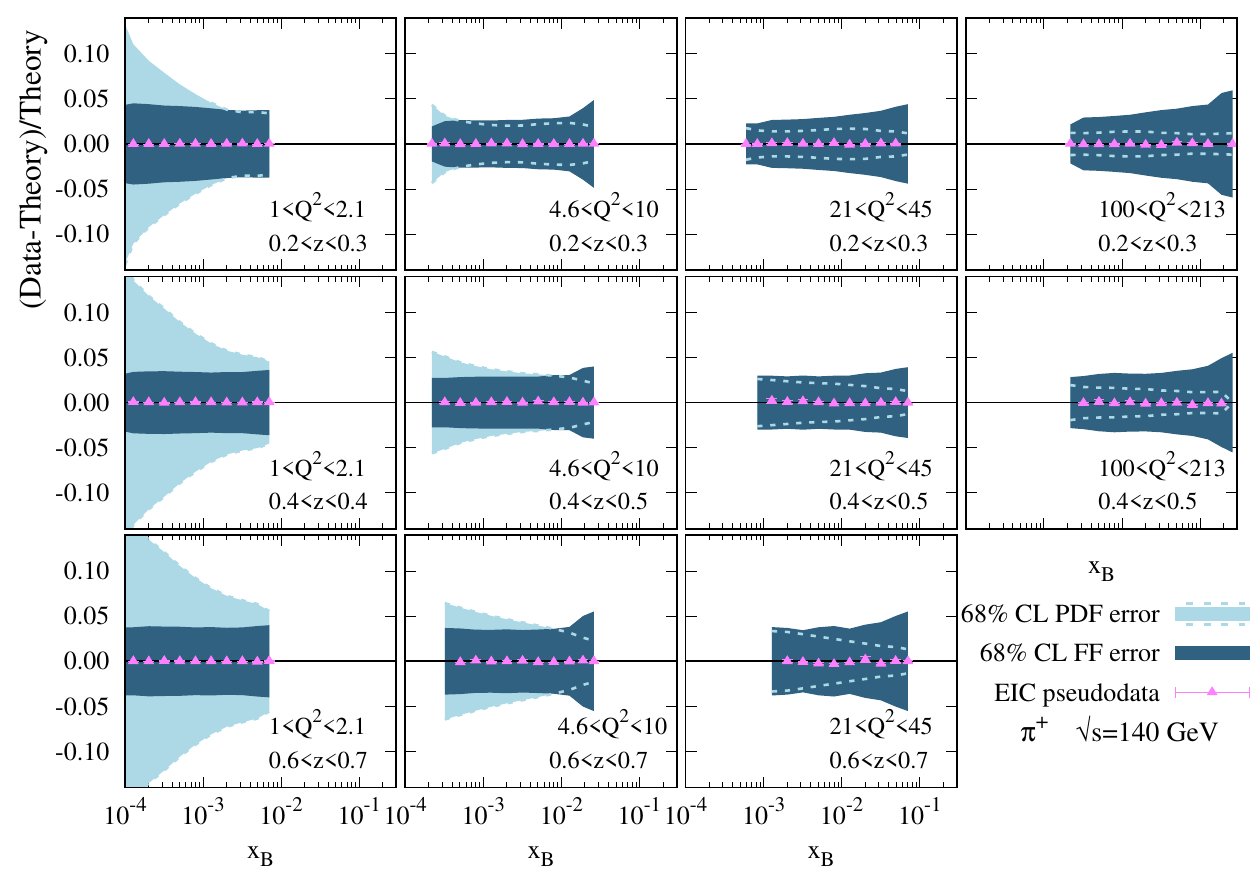,width=0.805\textwidth}
\caption{Pseudodata estimates for the production of $\pi^{+}$ at $\sqrt{s}=140$ GeV as a 
(Data-Theory)/Theory plot. The bands represent the uncertainty in the theoretical estimate 
coming from the PDFs (light blue) and from the FFs (dark-blue). The data is plotted for 
representative bins of $Q^{2}$ and $z$, as function of $x_{B}$. For those regions where the 
uncertainty coming form the FFs becomes larger than that of the PDFs, the error band of the latter 
is represented by the light blue dashed lines.\label{fig:data-theory}}
\end{figure*}

In Fig.~\ref{fig:data-theory}, we show the pseudodata estimates for the production of positively 
charged pions as a function of $x_{B}$ for representative bins of $Q^{2}$ and $z$. The pseudodata 
is presented in a (Data-Theory)/Theory plot together with the theoretical uncertainties for the 
cross section estimate coming from the PDFs (light-blue band) and from the FFs (dark-blue band). 
Clearly, while the uncertainties propagated from the FFs are roughly independent of $x_{B}$, 
those coming from the PDFs grow for smaller values of $x_{B}$, since at these values the PDFs 
are considerably less well-known than for the valence region. The FF uncertainty limits the 
impact of the reweighting process in the kinematic region where the PDF uncertainties are smaller. 
Iterating the procedure would yield smaller FF uncertainties. These in turn would constrain the 
PDFs better. In any case, we see from the first step of the iterative procedure shown here that 
the impact on the distributions is not negligible. Also, a combined PDF and FF global fit would 
circumvent these limitations. 

%
%
%
\begin{figure*}[bt!]
\vspace*{0.22cm}
\begin{center}
\hspace*{-0.1cm}
\epsfig{figure=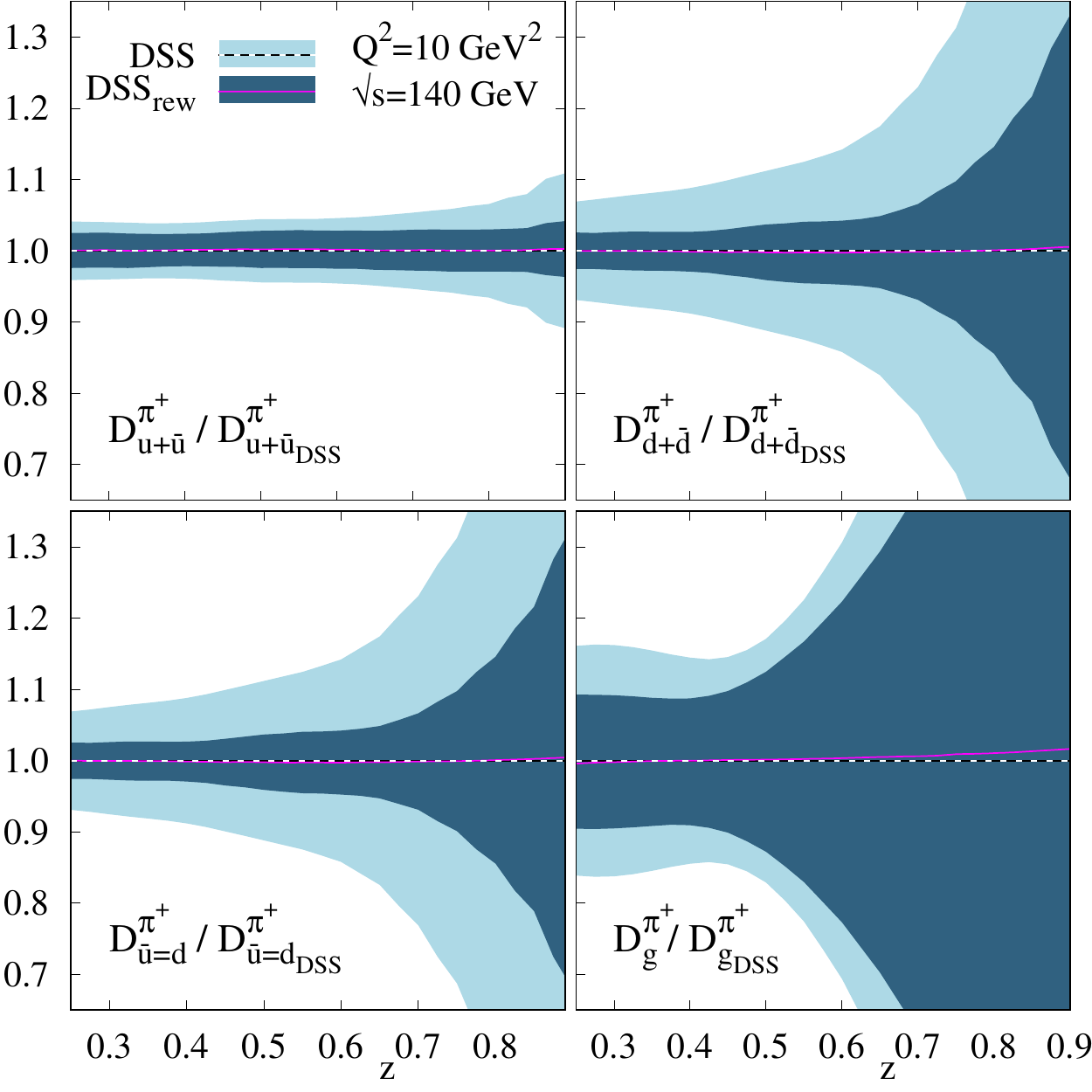,width=0.485\textwidth}
\epsfig{figure=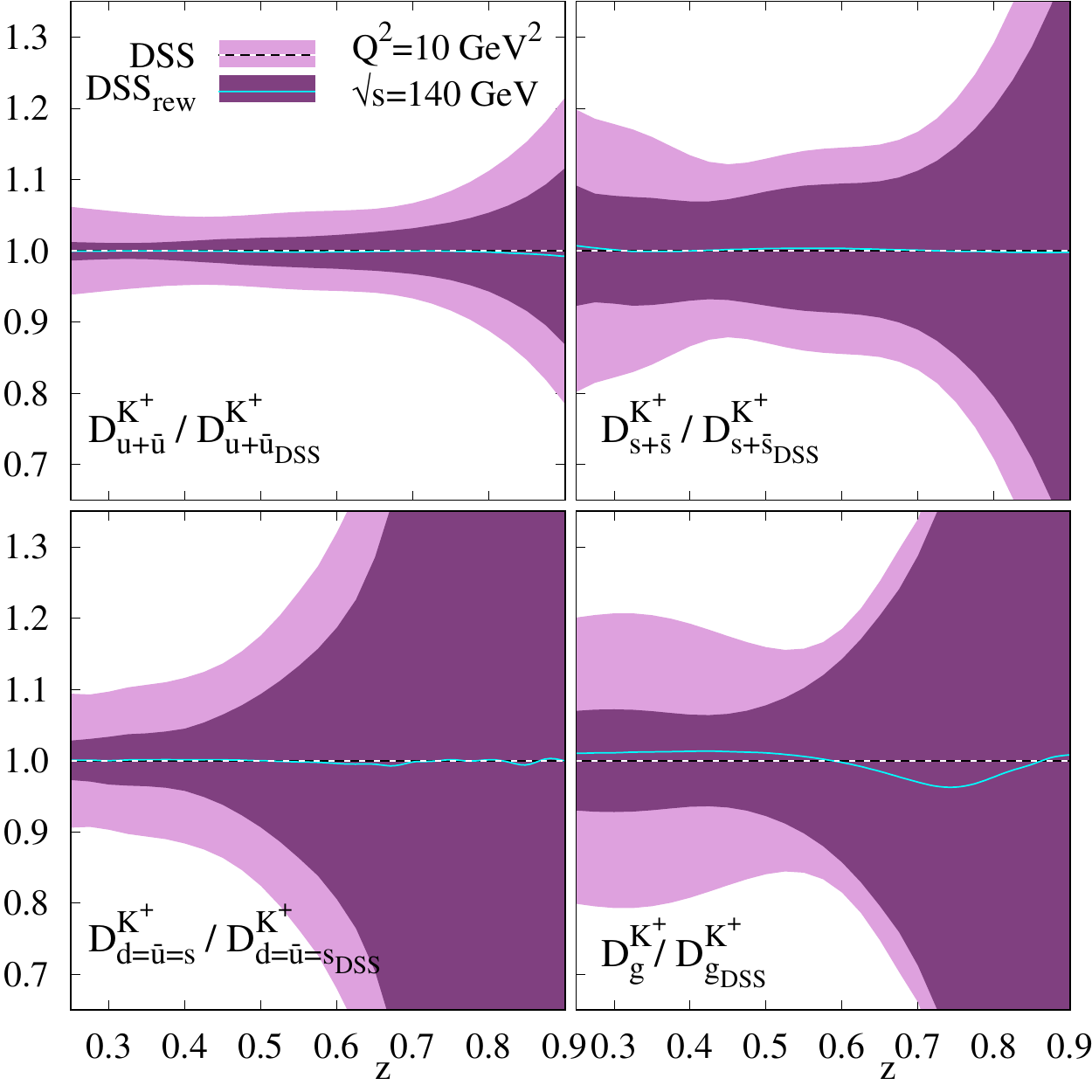,width=0.485\textwidth}
\end{center}
\vspace*{-0.5cm}
\caption{Reweighting of DSS NLO parton to pion and parton to kaon fragmentation-function replicas 
for the combinations $q+\bar{q}$ associated to the final hadron valence quarks (upper panels) as 
well as for the unfavoured flavours of quarks and gluons (lower panels) with EIC pseudodata of c.m.s energy $\sqrt{s}=140$ GeV. 
As in Fig.~\ref{fig:u+d}~and Fig.~\ref{fig:s+asymmetry}, the results are normalized to the DSS 
best fit. In the case of parton to pion FFs, the modified distributions are represented by the pink 
line, while their modified uncertainties are represented by the dark blue band. Analogously, the 
original central value and uncertainty are given by the black and white dashed line and the light 
blue band, respectively. The inverse color scheme is used in the case of parton to kaon FFs.
All results are shown at a scale of $Q^{2}=10$ GeV$^2$.
\label{fig:rew_ff_pion}}
\end{figure*} 
%

\bigskip
The results with pseudodata generated for the lower c.m.s. energy of $45\,\text{GeV}$, on 
the right-hand side, show that the reduction in the uncertainty bands is not as large as in 
the case of the higher c.m.s. energy. Nevertheless, the pseudodata for this configuration 
still imposes sizeable constraints on the distributions. The reweighting with this pseudodata
set leads to a reduction in the uncertainty of the order of $20\%$ in the case of the $u$ 
and $\bar{u}$ quarks, and around $10\%$ for the $d$ and $\bar{d}$ quark distributions. 
At variance with the higher c.m.s. energy, some deviations from the original best
fit are produced for $x_{B}<10^{-3}$, due to the absence of pseudodata points constraining 
the behavior of the replicas, which is fixed by the higher $x_{B}$ data. 

As for the higher c.m.s. energy, the kinematic region constrained by the inclusion of the new data 
at lower c.m.s. energy coincides with the region of larger values of the sensitivity coefficient, 
now restricted to $10^{-3}<x_{B}<10^{-2}$. Once again, it should be noticed that whereas the 
sensitivity coefficients suggest a more moderate impact for the higher $x_{B}$ region, the 
completely unmodified distributions are a result of the inclusion of the growing theoretical 
uncertainties coming from the FFs in the reweighting, which dilute the constraining power of 
the new data set.

Similar results are obtained for the (anti-)strange-quark distribution, which 
is depicted in Fig.~\ref{fig:s+asymmetry} (upper left panel), together with the flavour 
(upper right and lower left panels) and charge (lower right panel) symmetry breaking.
Again, the four panels on the left-hand side correspond to a set of pseudodata at
a c.m.s. energy of $\sqrt{s} = 140$ GeV, while those on the right-hand side, to the 
$\sqrt{s} = 45$ GeV set. As could be expected from the relatively poor determination 
of the strange-quark content of the proton, the most striking feature is an even more noticeable 
reduction in the uncertainty for the $s$ quark distribution, which is of the order of $75\%$ for 
momentum fractions below $10^{-2}$, driven by the kaon data through the reweighting.

The reduction in uncertainty of the strange-quark content of the proton has also a very
significant impact on the constraints for the so-called \textit{strange ratio}, shown in 
the upper left panel, which has been actively discussed in connection to recent LHC measurements. 
Our result indicates that EIC SIDIS data would be able to further constrain the $x_{B}$ dependence 
of the ratio, suggesting a rather asymmetric scenario at high $x_{B}$, while favouring 
SU(3) flavour symmetry between the light quarks for lower values of the momentum fraction.

Regarding the isospin and charge asymmetries, shown in the lower panels, no significant 
improvements in the uncertainty estimates are found. On the other hand, no important deviations 
from the original value are observed, which is fully consistent with the fact that the 
pseudodata was generated from theoretical estimates already containing the same degree of 
symmetry breaking, and the procedure does not introduce any spurious imbalance between 
$\bar{u}$ and $\bar{d}$ and between $s$ and $\bar{s}$.

The reweighting of the FF replicas yields comparable results in terms of impact, although 
with some specific features related to the FF extractions used as starting point.
In Fig.~\ref{fig:rew_ff_pion}, we show the effect of reweighting a set of $10^{5}$ replicas of 
the variants of the DSS14 and DSS17 sets of FFs (based on NNPDF3.0)
for pions and kaons with EIC SIDIS pseudodata for the c.m.s energy configuration of $\sqrt{s}=140$ GeV.
 In both cases, the sets of 
FF replicas are generated according to Eq.~\ref{eq:hessian_replicas}, from random variations 
in the parameter space, followed by an analogous application of the Bayesian inference 
procedure, described in previous sections. In both cases a sufficiently large number of 
effective replicas survives after the reweighting exercise, with $N_{eff}^{(\pi)}\approx 500$ 
and $N_{eff}^{(K)}\approx 200$.

As in the case of the PDF reweighting, the plots show the modified distributions and their 
estimated uncertainties normalized to the reference value of DSS FFs, depicted by the black 
and white dashed lines. The modified parton to pion FFs are represented by the solid (magenta) 
lines, and their uncertainties by the darker (blue) bands. The inverse color scheme is used 
with the parton to kaon FFs, with light-blue lines representing the modified FFs and violet 
bands representing their uncertainties. In both cases, the upper panels show  the FFs of 
the \textit{plus} combinations $D_{q+\bar{q}}^{H^{+}}$ associated to the
final hadron valence quarks, whereas the lower panels correspond to the FFs for the unfavoured 
light quarks and the gluon.  
Once again, since the pseudodata used for the reweighting procedure was generated smearing  
the NLO estimate with DSS sets of FFs, no important deviation from the original sets 
is to be expected.  

The improvement in the determination of both pion and kaon FFs is remarkable: 
In the case of parton to pion FFs, the reduction in the uncertainty of $D_{u+\bar{u}}^{\pi^{+}}$ is 
of order 25\%, while for $D_{d+\bar{d}}^{\pi^{+}}$, the reduction is of the order of 30\%.  
Even more impressive is the effect on the FFs associated to unfavored quark flavours, which show 
a reduction in the uncertainty of approximately 60\%.
This important improvement is mainly due to the  relatively poor constraints for the unfavoured 
flavors in the global fits. Notice that $D_{q}^{\pi^{+}}$ is assumed to be the same for 
$\bar{u}$ and $d$ in the global fit. 

It is also worth mentioning 
the impact of the pseudodata on the gluon to pion fragmentation function for low values of 
$z$, which shows a reduction of the uncertainty of the order of 40\%. In this case, the constraints 
come not only from the NLO contribution to the cross section associated to the hadronization 
of gluons, but also through the evolution equations, which depend critically on the gluon FF.  

As in the case of the reweighting with PDF replicas, the reweighting of FF replicas necessarily 
involves the inclusion of the theoretical uncertainties coming from the PDFs. Once again, the 
lack of variation of the distributions in the region of high $z$ is a result of the inclusion of 
uncertainties associated to the PDFs,  which grow with $z$ for a fixed value of $\{x_B,Q^{2}\}$, 
as can be seen in Fig.~\ref{fig:data-theory}. In the same sense, much of the expected impact of 
the low $Q^{2} $ data on $D_{u+\bar{u}}^{\pi^{+}}$ is diluted due to the growing uncertainty 
coming from the PDFs, resulting in a smaller impact.

Regarding the parton to kaon FFs, the results shown in Fig.~\ref{fig:rew_ff_pion} should 
be taken with some caution, since the much more rigid functional form assumed for 
some of the DSS kaon FFs could be too restrictive for the generation of faithful replicas.  
In fact, the reweighting results in a significantly lower number of effective replicas 
compared to the pion reweighting. 
While the constraints on the FFs for the combinations $u+\bar{u},s+\bar{s}$ are once again 
impressive, with reductions in the uncertainties of 
$D_{u+\bar{u}}^{K^{+}}$ and $D_{s+\bar{s}}^{K^{+}}$ around 70\% 
and 60\%, respectively, the less flexible parameterizations for the unfavoured FFs and gluons
could translate into an artificial reduction of the uncertainties. The comparison to actual
SIDIS data instead of simulated cross sections generated from the DSS sets will eventually 
indicate the need of a new FF fit with more flexibility or different flavour-symmetry assumptions.
In any case, the results clearly show that the EIC SIDIS measurements largely exceed in 
precision the current global analysis and therefore have a significant potential for the improvement 
of FF extractions. 

\section{Summary}\label{sec:summary}

The semi-inclusive production of hadrons in deep-inelastic electron-proton scattering
offers a remarkably versatile tool to probe both the flavour content of the proton 
and the way in which the different parton flavours confine into final-state
hadrons. QCD factorisation allows to model the corresponding cross sections in terms of 
non-perturbative parton distribution and fragmentation functions in such a way that 
precise cross-section measurements impose very stringent constraints on these distributions. 

The key advantage of SIDIS data in the determination of the PDFs lies in the fact 
that the flavour composition of the final-state hadrons probe a specific combination of 
partonic flavours, giving access to flavour-dependent information that is entangled  
in more inclusive measurements. Consequently, the unprecedented precision and kinematic 
coverage of SIDIS measurements at a future EIC will certainly enhance 
our knowledge on PDFs and FFs, and provide new insights into the inner structure of the 
nucleon, and the interactions among its most basic constituents. In this paper, we have
made quantitative assessment of the improvements. 

Despite the technical difficulties involved in a simultaneous extraction of both PDFs and 
FFs, techniques based on Bayesian inference allow to refine our knowledge on the 
non-perturbative distributions, including the critical information coming from SIDIS data. 
Through the implementation of reweighting techniques, we studied in detail the constraints 
that measurements at the future EIC would impose on both the parton distribution functions 
of the proton as well as on the parton to hadron fragmentation functions by using simulated 
data with realistic uncertainties.

We confirm the remarkable impact that EIC SIDIS data would have on the PDFs, especially 
on those of light quarks of radiative origin, which are comparatively less constrained than their 
valence counterparts. Our study suggests that outstanding reductions in the uncertainties 
of these distributions can be obtained, which we estimate to be of the order of $75\%$ in the 
case of the strange-quark content of the proton, $30\%$ for the up quark and $20\%$ for the down 
quark (for the most energetic configuration of $\sqrt{s} = 140 \text{GeV}$). In 
addition, our results indicate that it will be possible to constrain the strong parton
momentum fraction dependence of the \textit{strangeness ratio}, and have complementary 
estimates of the charge symmetry breaking.

We also find that the most significative effect on the parton distributions will be 
achieved with the much wider kinematic range covered by the EIC running at a large 
c.m.s. energy, for which more stringent constraints are found.

Regarding the fragmentation functions, we have also estimated the kinematic configurations 
where the EIC data could enhance the precision of FFs in future global analyses as well as the 
improvement in the precision of these distributions. Our results 
indicate that EIC SIDIS data would have a significant effect on the determination of the FFs, 
complementing the present measurements since they span a wider kinematic range than 
that currently probed. 

Our results highlight the importance that the forthcoming measurements at the EIC will 
have on the determination of the non-perturbative PDFs and FFs, taking them to a new 
standard in precision, and therefore refining our picture of the partonic structure of matter.
\bigskip
\section{Acknowledgments}

We warmly acknowledge Pia Zurita and Marco Stratmann for interesting comments and suggestions 
and their help with the reweighting methodologies. This work was supported in part by CONICET 
and ANPCyT.
CVH acknowledges the support from the Basque Government (Grant No. IT956-16) and the
Ministry of Economy and Competitiveness (MINECO) (Juan de la Cierva), Spain.
\clearpage



\begin{thebibliography}{99}
%
\bibitem{Butterworth:2015oua} 
  J.~Butterworth {\it et al.},
  J.\ Phys.\ G {\bf 43}, 023001 (2016)
  doi:10.1088/0954-3899/43/2/023001
  [arXiv:1510.03865 [hep-ph]].

\bibitem{Metz:2016swz} 
  A.~Metz and A.~Vossen,
  Prog.\ Part.\ Nucl.\ Phys.\  {\bf 91}, 136 (2016)
  doi:10.1016/j.ppnp.2016.08.003
  [arXiv:1607.02521 [hep-ex]].
  
  \cite{Rojo:2015acz}
\bibitem{Rojo:2015acz} 
  J.~Rojo {\it et al.},
  J.\ Phys.\ G {\bf 42}, 103103 (2015)
  doi:10.1088/0954-3899/42/10/103103
  [arXiv:1507.00556 [hep-ph]].
  
\bibitem{Rojo:2016ymp} 
  J.~Rojo,
  PoS DIS {\bf 2016}, 018 (2016)
  doi:10.22323/1.265.0018
  [arXiv:1606.08243 [hep-ph]].
  
  \bibitem{Feynman:1973xc} 
  R.~P.~Feynman,
  ``Photon-Hadron Interactions'', W.~A.~Benjamin, Inc. (Reading, Mass., 1972).
  
\bibitem{Field:1976ve} 
  R.~D.~Field and R.~P.~Feynman,
  Phys.\ Rev.\ D {\bf 15}, 2590 (1977).
  doi:10.1103/PhysRevD.15.2590
  
\bibitem{Sjo01} T. Sj\"{o}strand et al., Comp. Phys. Commun. {\bf 135} (2001) 238.
\bibitem{Sjo08} T. Sj\"{o}strand, S. Mrenna, P. Skands, Comp. Phys. Commun. {\bf 178} (2008) 852.

\bibitem{Borsa:2017vwy} 
  I.~Borsa, R.~Sassot and M.~Stratmann,
  Phys.\ Rev.\ D {\bf 96}, no. 9, 094020 (2017)
  doi:10.1103/PhysRevD.96.094020
  [arXiv:1708.01630 [hep-ph]].

\bibitem{Lees:2013rqd}
  J.~P.~Lees {\it et al.} [BaBar Collaboration],
  Phys.\ Rev.\ D {\bf 88}, 032011 (2013)
  doi:10.1103/PhysRevD.88.032011
  [arXiv:1306.2895 [hep-ex]].

\bibitem{Leitgab:2013qh}
  M.~Leitgab {\it et al.} [Belle Collaboration],
  Phys.\ Rev.\ Lett.\  {\bf 111}, 062002 (2013)
  doi:10.1103/PhysRevLett.111.062002
  [arXiv:1301.6183 [hep-ex]].

\bibitem{Airapetian:2012ki}
  A.~Airapetian {\it et al.} [HERMES Collaboration],
  Phys.\ Rev.\ D {\bf 87}, 074029 (2013)
  doi:10.1103/PhysRevD.87.074029
  [arXiv:1212.5407 [hep-ex]].

\bibitem{Adolph:2016bwc}
  C.~Adolph {\it et al.} [COMPASS Collaboration],
  Phys.\ Lett.\ B {\bf 767}, 133 (2017)
  doi:10.1016/j.physletb.2017.01.053
  [arXiv:1608.06760 [hep-ex]].

\bibitem{Agakishiev:2011dc}
  G.~Agakishiev {\it et al.} [STAR Collaboration],
  Phys.\ Rev.\ Lett.\  {\bf 108}, 072302 (2012)
  doi:10.1103/PhysRevLett.108.072302
  [arXiv:1110.0579 [nucl-ex]].

\bibitem{Abelev:2014laa}
  B.~B.~Abelev {\it et al.} [ALICE Collaboration],
  Phys.\ Lett.\ B {\bf 736}, 196 (2014)
  doi:10.1016/j.physletb.2014.07.011
  [arXiv:1401.1250 [nucl-ex]].
  
\bibitem{Accardi:2012qut} 
  A.~Accardi {\it et al.},
  Eur.\ Phys.\ J.\ A {\bf 52}, no. 9, 268 (2016)
  doi:10.1140/epja/i2016-16268-9
  [arXiv:1212.1701 [nucl-ex]].
  
\bibitem{Aschenauer:2014cki} 
  E.~C.~Aschenauer {\it et al.},
  arXiv:1409.1633 [physics.acc-ph].
  
\bibitem{Aschenauer:2015ata} 
  E.~C.~Aschenauer, R.~Sassot and M.~Stratmann,
  Phys.\ Rev.\ D {\bf 92}, no. 9, 094030 (2015)
  doi:10.1103/PhysRevD.92.094030
  [arXiv:1509.06489 [hep-ph]].
  
\bibitem{Aschenauer:2016our} 
  E.~C.~Aschenauer {\it et al.},
  arXiv:1602.03922 [nucl-ex].
  
\bibitem{Ball:2010gb} 
  R.~D.~Ball {\it et al.} [NNPDF Collaboration],
  Nucl.\ Phys.\ B {\bf 849}, 112 (2011)
  Erratum: [Nucl.\ Phys.\ B {\bf 854}, 926 (2012)]
  Erratum: [Nucl.\ Phys.\ B {\bf 855}, 927 (2012)]
  doi:10.1016/j.nuclphysb.2011.03.017, 10.1016/j.nuclphysb.2011.10.024, 10.1016/j.nuclphysb.2011.09.011
  [arXiv:1012.0836 [hep-ph]].
  
\bibitem{Ball:2011gg} 
  R.~D.~Ball {\it et al.},
  Nucl.\ Phys.\ B {\bf 855}, 608 (2012)
  doi:10.1016/j.nuclphysb.2011.10.018
  [arXiv:1108.1758 [hep-ph]].
  
\bibitem{Armesto:2013kqa} 
  N.~Armesto, J.~Rojo, C.~A.~Salgado and P.~Zurita,
  JHEP {\bf 1311}, 015 (2013)
  doi:10.1007/JHEP11(2013)015
  [arXiv:1309.5371 [hep-ph]].
  
\bibitem{Paukkunen:2014zia} 
  H.~Paukkunen and P.~Zurita,
  JHEP {\bf 1412}, 100 (2014)
  doi:10.1007/JHEP12(2014)100
  [arXiv:1402.6623 [hep-ph]].
  
\bibitem{Bertone:2018ecm} 
  V.~Bertone {\it et al.} [NNPDF Collaboration],
  Eur.\ Phys.\ J.\ C {\bf 78}, no. 8, 651 (2018)
  doi:10.1140/epjc/s10052-018-6130-4
  [arXiv:1807.03310 [hep-ph]].
  
\bibitem{Wang:2018heo} 
  B.~T.~Wang, T.~J.~Hobbs, S.~Doyle, J.~Gao, T.~J.~Hou, P.~M.~Nadolsky and F.~I.~Olness,
  arXiv:1803.02777 [hep-ph].
  
  
\bibitem{Furmanski:1981cw} 
  W.~Furmanski and R.~Petronzio,
  Z.\ Phys.\ C {\bf 11}, 293 (1982).
  doi:10.1007/BF01578280
  
\bibitem{Graudenz:1994dq} 
  D.~Graudenz,
  Nucl.\ Phys.\ B {\bf 432}, 351 (1994)
  doi:10.1016/0550-3213(94)90606-8
  [hep-ph/9406274].
  
\bibitem{deFlorian:1997zj} 
  D.~de Florian, M.~Stratmann and W.~Vogelsang,
  Phys.\ Rev.\ D {\bf 57}, 5811 (1998)
  doi:10.1103/PhysRevD.57.5811
  [hep-ph/9711387].
  
  
\bibitem{deFlorian:2014xna} 
  D.~de Florian, R.~Sassot, M.~Epele, R.~J.~Hernández-Pinto and M.~Stratmann,
  Phys.\ Rev.\ D {\bf 91}, no. 1, 014035 (2015)
  doi:10.1103/PhysRevD.91.014035
  [arXiv:1410.6027 [hep-ph]].
  
\bibitem{deFlorian:2017lwf} 
  D.~de Florian, M.~Epele, R.~J.~Hernandez-Pinto, R.~Sassot and M.~Stratmann,
  Phys.\ Rev.\ D {\bf 95}, no. 9, 094019 (2017)
  doi:10.1103/PhysRevD.95.094019
  [arXiv:1702.06353 [hep-ph]].
  
\bibitem{deFlorian:2007aj} 
  D.~de Florian, R.~Sassot and M.~Stratmann,
  Phys.\ Rev.\ D {\bf 75}, 114010 (2007)
  doi:10.1103/PhysRevD.75.114010
  [hep-ph/0703242 [HEP-PH]].


\bibitem{Daleo:2004pn} 
  A.~Daleo, D.~de Florian and R.~Sassot,
  Phys.\ Rev.\ D {\bf 71}, 034013 (2005)
  doi:10.1103/PhysRevD.71.034013
  [hep-ph/0411212].
  

\bibitem{Daleo:2003xg} 
  A.~Daleo, C.~A.~Garcia Canal and R.~Sassot,
  Nucl.\ Phys.\ B {\bf 662}, 334 (2003)
  doi:10.1016/S0550-3213(03)00334-1
  [hep-ph/0303199].

\bibitem{Daleo:2003jf} 
  A.~Daleo and R.~Sassot,
  Nucl.\ Phys.\ B {\bf 673}, 357 (2003)
  doi:10.1016/j.nuclphysb.2003.09.007
  [hep-ph/0309073].
 
\bibitem{Ball:2014uwa} 
  R.~D.~Ball {\it et al.} [NNPDF Collaboration],
  JHEP {\bf 1504}, 040 (2015)
  doi:10.1007/JHEP04(2015)040
  [arXiv:1410.8849 [hep-ph]].
  
 
\bibitem{Nadolsky:2008zw} 
  P.~M.~Nadolsky, H.~L.~Lai, Q.~H.~Cao, J.~Huston, J.~Pumplin, D.~Stump, W.~K.~Tung and C.-P.~Yuan,
  Phys.\ Rev.\ D {\bf 78}, 013004 (2008)
  doi:10.1103/PhysRevD.78.013004
  [arXiv:0802.0007 [hep-ph]].
  
\bibitem{Ball:2008by} 
  R.~D.~Ball {\it et al.} [NNPDF Collaboration],
  Nucl.\ Phys.\ B {\bf 809}, 1 (2009)
  Erratum: [Nucl.\ Phys.\ B {\bf 816}, 293 (2009)]
  doi:10.1016/j.nuclphysb.2008.09.037, 10.1016/j.nuclphysb.2009.02.027
  [arXiv:0808.1231 [hep-ph]].
  
\bibitem{Guffanti:2010yu} 
  A.~Guffanti and J.~Rojo,
  Nuovo Cim.\ C {\bf 033}, no. 4, 65 (2010)
  doi:10.1393/ncc/i2010-10668-y
  [arXiv:1008.4671 [hep-ph]].
 
  
  
  \end{thebibliography}
\end{document}